\documentclass[aps, prl, superscriptaddress, preprintnumbers, twocolumn, floatfix, longbibliography,10pt]{revtex4-2}

\usepackage{soul}
\usepackage{mathdots}
\usepackage{graphicx, xcolor}
\usepackage{amsfonts,amsmath,amssymb,amstext,dsfont,bm}
\usepackage{float,wrapfig,subfigure, psfrag}
\usepackage{comment}

\usepackage[colorlinks=true,urlcolor=blue,citecolor=blue,linkcolor=blue, breaklinks=true]{hyperref}

\def\bra#1{\mathinner{\langle{#1}|}}
\def\ket#1{\mathinner{|{#1}\rangle}}

\newcommand{\be}{\begin{equation}}
\newcommand{\ee}{\end{equation}}

\newcommand{\bea}{\begin{eqnarray}}
\newcommand{\eea}{\end{eqnarray}}

\newcommand{\bk}{\mathbf{k}}

\definecolor{purple}{rgb}{0.8,0,0.6}
\definecolor{darkgreen}{rgb}{0.00,0.6,0.00}

\definecolor{Green}{rgb}{0,0.55,0}

\extrafloats{100}

\begin{document}

\title{Dynamical unmasking of $p$-wave magnetism by ac spin and dc orbital polarizations}

\author{Yantao Li}
\affiliation{Department of Physics and Astronomy, University of Missouri, Columbia, Missouri, 65211, USA}

\author{Jacob Linder}
\affiliation{Center for Quantum Spintronics, Department of Physics, Norwegian University of Science and Technology, NO-7491 Trondheim, Norway}

\author{Pavlo~Sukhachov}
\email{pavlo.sukhachov@missouri.edu}
\affiliation{Department of Physics and Astronomy, University of Missouri, Columbia, Missouri, 65211, USA}
\affiliation{MU Materials Science \& Engineering Institute,
University of Missouri, Columbia, Missouri, 65211, USA}

\date{September 14, 2026}

\begin{abstract}
Odd-parity magnets have strong spin polarization in momentum space but no net magnetization, making their spin texture hidden from conventional magnetometry. We show that a monochromatic electric field dynamically lifts this cancellation in a $p$-wave magnet. Within each drive cycle, the field produces a net ac spin polarization linear in the driving field whose interband component is resonantly enhanced at the nonrelativistic exchange gap. Because this polarization oscillates at the drive frequency, it vanishes upon period averaging. The orbital channel follows a different hierarchy: its linear response cancels in momentum space, whereas interband coherence produces a rectified dc orbital polarization at second order when the field has components parallel and transverse to the $p$-wave spin-splitting direction. 
Together, the ac spin and dc orbital responses dynamically unmask $p$-wave magnetism without requiring any spin-orbit coupling.
\end{abstract}

\maketitle

\indent\textcolor{blue}{\em Introduction}---
Odd-parity magnets are unconventional magnets characterized by momentum-odd spin polarization of their electronic bands~\cite{Hellenes-Smejkal:2023-v3, Brekke-Linder-MinimalModelsTransport-2024, Yu-Agterberg-OddParityMagnetismDriven-2025, Song-Comin-ElectricalSwitchingPwave-2025, Kunes-Geffroy-SpontaneousSpinTextures-2016, Lin-Vila-OddparityAltermagnetismSublattice-2026, Zhuang-Yan-OddParityAltermagnetismOriginated-2025, Leeb-Knolle-CollinearpwaveMagnetism-2026}. Because the spin texture changes sign under momentum inversion, its integrated value vanishes in equilibrium. Experimental evidence for odd-parity magnetism include Gd$_3$(Ru$_{1-\delta}$Rh$_\delta$)$_4$Al$_{12}$~\cite{Yamada-Hirschberger-MetallicPwaveMagnet-2025}, NiI$_2$~\cite{Song-Comin-ElectricalSwitchingPwave-2025}, and CeNiAsO~\cite{Zhang-Chen-OddSpinSymmetry-2026}. Among odd-parity magnets, $p$-wave magnets are the simplest realization, which relies on the combined translation and time-reversal symmetry (TRS) $\bm{\tau}\mathcal T$~\cite{Hellenes-Smejkal:2023-v3, Brekke-Linder-MinimalModelsTransport-2024, Luo-Law-SpinGroupSymmetry-2026, Mitscherling-Smejkal-MicroscopicOrigin$p$wave-2026, li2026pwaveorbitalmagnetism}.
This symmetry enforces vanishing net magnetization and, for certain structures, protects the odd-parity splitting from the effects of spin-orbit coupling (SOC)~\cite{Kudasov:2024, Hodt-Linder-FateWaveSpin-2025}.
Beyond $p$-wave magnets, odd-parity textures with $f$- and $h$-wave symmetry are also possible~\cite{Hayami-Kusunose-SpontaneousAntisymmetricSpin-2020, Hayami-Kusunose-BottomupDesignSpinsplit-2020, Yu-Agterberg-OddParityMagnetismDriven-2025, Luo-Law-SpinGroupSymmetry-2026, Hirschmann-Hirschberger-SymmetryEnforcedNodalfWave-2026}. The momentum-space structure of odd-parity magnets gives rise to a broad range of phenomena, including anisotropic transport~\cite{Brekke-Linder-MinimalModelsTransport-2024, Hellenes-Smejkal:2023-v3, Hedayati-Salehi-TransverseSpinCurrent-2025, Ezawa-ThirdorderFifthorderNonlinear-2025, Zhou-Li-AnisotropicResistivity$p$wave-2025, Ezawa-TunnelingMagnetoresistanceJunction-2026}, superconducting phenomena~\cite{Sukhachov-Linder-CoexistencePwaveMagnetism-2025, Maeda-Tanaka:2024, Kokkeler-Bergeret-QuantumTransportTheory-2025, Maeda-Cayao-ClassificationPairSymmetries-2025, Nagae-Ikegaya-MajoranaFlatBands-2025, Fukaya-Tanaka-TunnelingConductanceSuperconducting-2025, Bobkov-Bobkova-ProximityEffectWave-2025, Sun-Law-PseudoIsingSuperconductivityInduced-2025, Fukaya-Tanaka-pwaveSuperconductivityJosephson-2026, Froldi-Freire-HighlyEfficientSuperconducting-2026, Khodas-Mazin-NonrelativisticIsingSuperconductivityPwave-2026, Pal-Saha-EmergentSuperconductingPhases-2026}, and photogalvanic effects~\cite{Song-Comin-ElectricalSwitchingPwave-2025, Cuono-Picozzi-ChargeSpinPhotogalvanic-2026}.  

Odd-parity magnets pose a distinctive readout problem directly connected to their defining property: their electronic bands can be strongly spin polarized, but being odd in momentum, they lead to vanishing equilibrium net magnetization, denying conventional magnetometry techniques.
An electric field can reveal a momentum-odd spin texture by changing the electronic distribution in momentum space. The resulting generation of a net spin polarization is known as the spin Edelstein effect (SEE). In odd-parity magnets, this response is governed by nonrelativistic exchange scales rather than weak relativistic SOC~\cite{Chakraborty-Sinova-HighlyEfficientNonrelativistic-2024, Ezawa-OutofplaneEdelsteinEffects-2025, Kim-Park-OddParityMagnetismGateTunable-2026, Sim-Rachel-QuantumSpinModels-2026, Hirschmann-Hirschberger-SymmetryEnforcedNodalfWave-2026}. Its orbital counterpart, the orbital Edelstein effect (OEE), converts an electric field into orbital angular momentum~\cite{Levitov-Nazarov-MagnetoelectricEffectsConductors-1985, Yoda-Murakami-CurrentinducedOrbitalSpin-2015, Yoda-Murakami-OrbitalEdelsteinEffect-2018}. 
Although static Edelstein responses in unconventional magnets have attracted considerable attention, their time-dependent spin and nonlinear orbital responses as well as intraperiod dynamics remain less understood. 

Floquet theory provides an efficient way to study this regime and address these questions, describing a periodically driven system in terms of quasienergy bands and an intraperiod micromotion~\cite{Sam73a, Goldman-Dalibard-PeriodicallydrivenQuantumSystems-2014, Bukov-Polkovnikov-UniversalHighFrequencyBehavior-2015, Eckardt-Anisimovas-HighfrequencyApproximationPeriodically-2015, Eckardt-ColloquiumAtomicQuantum-2017, Oka-Kitamura-FloquetEngineeringQuantum-2019, Mori-FloquetStatesOpen-2023, Sato-Ikeda-FloquetTheoryApplications-2025}. Floquet engineering has been used to modify the magnetic and electronic properties in altermagnets~\cite{yarmohammadiAnisotropicLighttailoredRKKY2025, Fu-Cayao-LightinducedFloquetSpintriplet-2025, Fu-Cayao-FloquetEngineeringSpin-2026, Yokoyama-FloquetEngineeringTriplet-2025, Yarmohammadi-Oppeneer-SpinPolarizationEngineering-2026, Li-Kovalev-FloquetSpinSplitting-2026, Yarmohammadi-Freericks-FloquetinducedAnisotropicMagnetoresistance-2026} and odd-parity magnets~\cite{Fu-Cayao-LightinducedFloquetSpintriplet-2025, yu2026tunableoddparityspinsplittings, Fu-Cayao-FloquetEngineeringSpin-2026, Huang-Wang-LightinducedOddparityMagnetism-2025, Zhu-Ruan-FloquetOddparityCollinear-2025, Liu-Yan-LightinducedOddparityAltermagnets-2025, Yarmohammadi-Oppeneer-GiantPerpendicularEdelstein-2026}, respectively.
The possibility of converting even-parity into odd-parity spin texture via a two-color driving field was mentioned in Ref.~\cite{yu2026tunableoddparityspinsplittings}.

In this work, we show that a monochromatic electric field can dynamically unmask $p$-wave magnetism by generating two complementary responses: ac spin and rectified dc orbital responses.
Within each drive cycle, the spin texture acquires an even-in-momentum component, resulting in a nonzero net ac spin polarization oscillating at the drive frequency and linear in the field. This polarization vanishes upon period averaging. Its interband contribution is resonantly enhanced near the minimum of the direct gap determined by the nonrelativistic exchange, rather than SOC. The resonance originates from spin precession generated by the noncommuting spin-splitting and exchange terms in the Hamiltonian.

The orbital response follows a different hierarchy. At linear order, the induced orbital texture is momentum-odd and, therefore, vanishes after integrating over momentum. At second order, a field with components both parallel and transverse to the spin-splitting direction generates a momentum-even rectified orbital polarization. This response originates from interband coherence encoded in the Berry-connection matrix elements entering the orbital magnetic moment. Together, these ac spin and dc orbital responses provide a SOC-free route to dynamically probe $p$-wave magnetism.

\begin{figure}[t]
\centering
\includegraphics[width=0.48\textwidth]{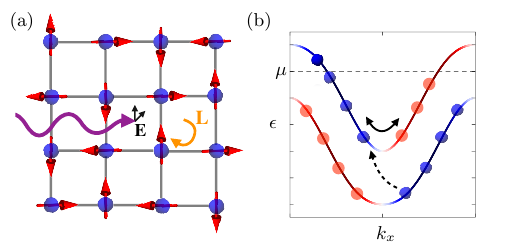}
\caption{Panel (a) shows a schematic of a light-driven $p$-wave magnet. Light, via an electric field $\mathbf{E}$, induces oscillating spin polarization and rectified orbital angular momentum $\mathbf{L}$. Panel (b) illustrates the spin Edelstein effect, where the electric field redistributes carriers within the spin-polarized bands (solid arrows), while the dashed arrow denotes an interband transition responsible for the resonant response near the minimum direct gap. Red and blue indicate opposite spin polarizations.
}
\label{fig:cover}
\end{figure}

\begin{figure*}[t]
   \centering
   \includegraphics[width=7.in]{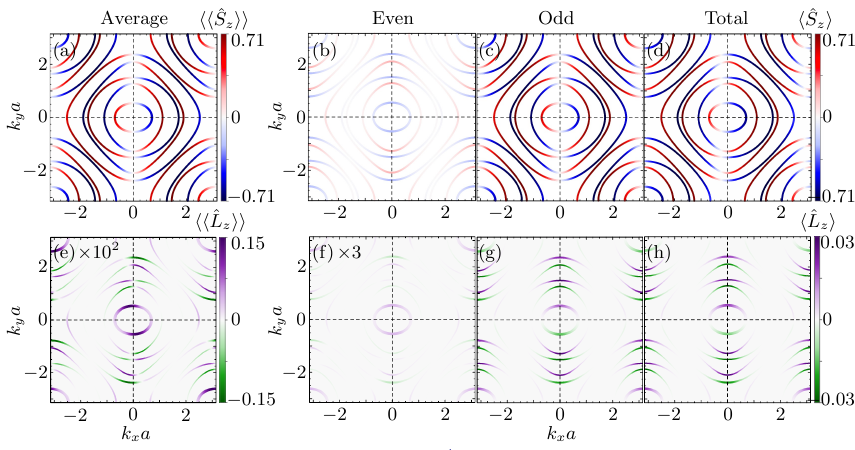}
   \caption{Panels (a) and (e) show the period-averaged spin and orbital angular momentum textures, respectively. Panels (b)–(d) and (f)–(h) show the instantaneous spin and orbital angular momentum textures at $t=0.5\,T_d$ in momentum space. The even, odd, and total contributions are arranged from left to right. In all panels, we sum over two sectors. The parameters in the model \eqref{eq:Ht} are set to $eA_x=eA_y=0.1\,a^{-1}$, $J_{sd}=\tilde{\alpha}_x=\lambda$, $\tilde{\alpha}_y=0$, $\omega = 2\pi J_{sd}/4$, and $\varphi=0$.
   We multiplied the results in panels (e) and (f) by $100$ and $3$, respectively. 
   }
   \label{fig:micromotion}
\end{figure*}

\indent\textcolor{blue}{\em Model}---
To demonstrate the nontrivial dynamical response of odd-parity magnets, we use the following minimal lattice model for a $p$-wave magnet~\cite{Brekke-Linder-MinimalModelsTransport-2024}
\begin{eqnarray}
\label{eq:Ht}
H(\bk) &=&  \xi_{\mathbf{k}}\,\sigma_0\otimes\tau_0 + M(\mathbf{k})\,\sigma_z\otimes\tau_0 + J_{sd}\,\sigma_x\otimes\tau_z \nonumber\\
&=& -2\lambda[\cos\!\big(k_x a\big)
+ \cos\!\big(k_y a \big)]
\sigma_0\otimes\tau_0 \nonumber\\
&+& \sum_{i=x,y} \tilde{\alpha}_i\sin{(k_i a)} \sigma_z\otimes\tau_0+ J_{sd}\,\sigma_x\otimes\tau_z.
\end{eqnarray}
Here, $\lambda$ is the nearest-neighbor hopping amplitude, $\tilde{\bm{\alpha}}$ is the spin-splitting vector, and $J_{sd}$ is the $sd$ coupling strength. The Pauli matrices $\bm{\sigma}$ and $\bm{\tau}$ act in the spin and sector spaces~\cite{Brekke-Linder-MinimalModelsTransport-2024}, respectively. 

The exchange term $J_{sd}\sigma_x\otimes\tau_z$ breaks ordinary TRS~\footnote{The importance of the TRS breaking was also emphasized in a different context in Ref.~\cite{Fukaya-Tanaka-TunnelingConductanceSuperconducting-2025}.} in each sector. However, the two sectors are exchanged by $\tau_x$, and the full Hamiltonian preserves the combined antiunitary symmetry $\tau_x\mathcal{T}$. This symmetry relates states at $\mathbf{k}$ and $-\mathbf{k}$ with opposite spin polarizations and is crucial for momentum-odd spin texture. Furthermore, the exchange term does not commute with the spin splitting term $M(\mathbf{k})\,\sigma_z\otimes\tau_0$, which, as we will show below, allows for spin precession and interband coherence.

To make an analytical advance, we also use a continuum model derived by expanding the Hamiltonian \eqref{eq:Ht} around the $\Gamma$ point and projecting onto a sector $\eta=\pm$,
\begin{equation}
\label{ate-H-def}
H(\bk) = \xi_{\bk} \sigma_0 +M(\bk) \sigma_z +\eta J_{sd} \sigma_x,
\end{equation}
where $\xi_{\bk}=k^2/(2m)$ is the kinetic energy with $(2m)^{-1}=\lambda a^2$, $M(\bk)=(\bm{\alpha}\cdot\bk)$ is the spin-splitting term with $a\tilde{\bm{\alpha}} = \bm{\alpha}$, and we absorbed the constant energy shift $-4\lambda$. Notice that the Hamiltonian \eqref{ate-H-def} has the form of a Hamiltonian in an effective momentum-dependent Zeeman field $\mathbf{B}(\bk) = \left\{\eta J_{sd},0,M(\bk)\right\}$. In the presence of an electromagnetic field, we perform the standard substitution $\bk \to \bk +e \mathbf{A}(t)$, where $-e<0$ is the charge of the electron. Here and in what follows, we set $\hbar =c=1$.

\indent\textcolor{blue}{\em Floquet approach and Micromotion}---
For a periodically driven system with $H(t+T_d)=H(t)$, we use Floquet theory to describe the rearrangement of the Floquet energy bands and the resulting change of the spin and orbital angular momentum polarizations. 

We consider the periodic vector potential $\mathbf{A}(t)=\left\{A_x\cos(\omega t),\,A_y\cos(\omega t +\varphi)\right\}$ with $\varphi$ controlling the light polarization and $\omega = 2\pi/T_d$ being the light frequency. Then, the Floquet quasienergies $\varepsilon_{\mathbf{k},n}$ are defined modulo the driving frequency $\omega$ and are taken in the first Floquet Brillouin zone $(-\omega/2,\omega/2]$. 
To provide information about the behavior of the driven system on the subperiod timescale, we use the Floquet micromotion approach. The Floquet formalism is summarized in the End Matter, with further details provided in the Supplemental Material (SM)~\cite{SM}. 

The Floquet micromotion results for the mean values of the $z$-components of the spin operator $\hat{S}_z = \sigma_z\otimes \tau_0/2$ and the orbital angular momentum operator $\hat{L}_z = -1/(g_L \mu_B) \hat{m}_z$~\footnote{In the definition of the orbital angular momentum, $g_L=1$ is the Land\'{e} $g$-factor and $\mu_B = e/(2m_e)$ is the Bohr magneton.} are shown in Fig.~\ref{fig:micromotion}. During the micromotion, the spin texture develops a momentum-even component, which shifts the nodal line of the total texture away from the $\Gamma$ point, as shown in Figs.~\ref{fig:micromotion}(b)--\ref{fig:micromotion}(d). This component oscillates during the driving cycle and vanishes after period averaging, leaving a momentum-odd spin texture. The orbital angular momentum also develops momentum-even and momentum-odd instantaneous components. However, in contrast to the spin response, its period-averaged texture contains only a momentum-even component. Its quadratic dependence on the driving amplitude indicates that it originates from a second-order rectification process~\cite{SM}. 

\indent\textcolor{blue}{\em Dynamical spin response}---
To clarify the microscopic origin of the momentum-even spin polarization observed during the micromotion, we consider the band-resolved spin dynamics in the continuum model. Within a single-relaxation-time approximation and for a fixed sector $\eta$, the spin expectation value obeys the following equation:
\begin{equation}
\label{ate-dis1-S-eff-1}
\partial_t \mathbf{S}(\bk,t) = 2\, \mathbf{B}(\bk,t)\times\mathbf{S}(\bk,t) -\frac{\mathbf{S}(\bk,t) -\mathbf{S}_0(\bk)}{\tau},
\end{equation}
where $\tau$ is the relaxation time~\footnote{While being simple, the relaxation time approach does not distinguish between Dyakonov-Perel and Elliott-Yafet mechanisms.}. The equilibrium spin polarization $\mathbf{S}_{0}(\bk)$ reads $\bm{S}_{0}(\bk) = s \mathbf{B}(\mathbf{k})/[2B(\mathbf{k})]$, where $s=\pm$ corresponds to the bands $\epsilon_{s} = \xi_k +s B(\mathbf{k})$ of the Hamiltonian in Eq.~\eqref{ate-H-def}. Note that $S_{z,0}(\bk) \propto M(\mathbf{k})$ and is odd in momentum.

We linearize Eq.~\eqref{ate-dis1-S-eff-1} in the electric field $\mathbf{E}(t)=\mathbf{E}_\omega e^{-i\omega t}+{\rm c.c.}$ and write $S_z(t)=S_{z,0}+S_{z,\omega}e^{-i\omega t}+{\rm c.c.}$ The sector-summed response is
\begin{eqnarray}
S_{z,\omega}(\mathbf{k}) &=&  \frac{8J_{sd}^2S_{z,0}}{i\omega M(\mathbf{k})} \frac{ e\mathbf{E}_\omega\cdot\boldsymbol{\alpha}}{D_\omega(\mathbf{k})}
=s \frac{4J_{sd}^2}{i\omega B(\mathbf{k})} \frac{ e\mathbf{E}_\omega\cdot\boldsymbol{\alpha}}{D_\omega(\mathbf{k})},
\label{ate-dis1-tau-app-Sz-harmonic-omega}
\end{eqnarray}
where $D_\omega(\mathbf{k}) = \gamma_\omega^2+4B^2(\mathbf{k})$ and $\gamma_\omega=\tau^{-1}-i\omega$. The induced $x$ and $y$ components of $\mathbf{S}(\mathbf{k})$ cancel after summing over $\eta$. The solution in Eq.~\eqref{ate-dis1-tau-app-Sz-harmonic-omega} is momentum-even. The origin of this even component of the spin polarization can be traced to the TRS-breaking $J_{sd}$ term that couples different spin projections and allows the nontrivial spin polarization components $S_{x,y}$ in each of the sectors to result in a momentum-even $S_{z}$, see SM~\cite{SM} for details. The denominator in Eq.~\eqref{ate-dis1-tau-app-Sz-harmonic-omega} acquires minimal value near $\omega=2B(\mathbf{k})$, corresponding to the resonant spin precession between the two bands. In materials with Rashba or Dresselhaus SOC, 
such a resonance is set by the SOC-induced splitting rather than by the nonrelativistic exchange gap
~\cite{Saleh-Maiti-ResonantEdelsteinInverse-2026}. 

The spin dynamics equation \eqref{ate-dis1-S-eff-1} establishes the spin texture in the momentum space and shows resonance but does not determine the occupation-weighted total spin polarization. Therefore, we calculate the dynamical spin Edelstein response,
\begin{equation}
M_i^{(S)} = \chi_{ij}^{(S)}(\omega)E_j(\omega), \quad \chi_{ij}^{(S)} = \chi_{ij}^{(S);{\rm intra}} + \chi_{ij}^{(S);{\rm inter}},
\end{equation}
using the Kubo formalism. The inter- and intraband parts of the spin Edelstein response tensor are
\begin{eqnarray}
\label{kubo-chiO-1}
\chi_{ij}^{(S); {\rm inter}}(\omega) &=& -i e \sum_{a\neq b} \int \frac{d^2k}{(2\pi)^2} \frac{n_F(\epsilon_b) -n_F(\epsilon_a)}{\epsilon_b -\epsilon_a} \nonumber\\
&\times&\frac{S_{i;ab} v_{j;ba}}{\omega +i/\tau -\epsilon_b +\epsilon_a}
\end{eqnarray}
\begin{equation}
\label{kubo-chiO-intra}
\chi_{ij}^{(S); {\rm intra}}(\omega) = \frac{i}{\omega +i/\tau} e \sum_a \int \frac{d^2k}{(2\pi)^2} n_F'(\epsilon_a) S_{i;aa} v_{j;aa},
\end{equation}
respectively.
Here, the summation $\sum_a$ includes the summation over bands $\epsilon_{\pm}$ and sectors $\eta=\pm$. The matrix element of the velocity operator, $\hat{v}_i = \partial_{k_i}H(\bk)$, and the spin operator, $\hat{S}_i$, are denoted as $v_{i;ab}$ and $S_{i;ab}$, respectively. The Fermi-Dirac distribution is denoted by $n_{F} (\epsilon_a) = 1/\left[1 + e^{(\epsilon_a -\mu)/T}\right]$, where $\mu$ is the chemical potential and prime denotes the derivative with respect to the argument. See the End Matter and SM~\cite{SM} for details. 

At vanishing temperature, for $\mu>J_{sd}$, and to the leading order in $\alpha k_F/J_{sd}$, the nontrivial response-tensor component are 
\begin{eqnarray}
\label{kubo-chiO-intra-1-small-alpha}
\chi_{zj}^{(S); {\rm intra}}(\omega) &\approx& \frac{em\alpha_j \tau}{\pi} \frac{1 +i\omega \tau}{1 +(\omega \tau)^2},
\end{eqnarray} 
\begin{eqnarray}
\label{kubo-chiO-inter-app-1}
\chi_{zj}^{(S); {\rm inter}}(\omega) &\approx& -i \frac{em \alpha_j}{\pi} \frac{\omega +i/\tau}{(\omega +i/\tau)^2 -4J_{sd}^2}.
\end{eqnarray} 

\begin{figure}[t]
\centering
\includegraphics[width=0.48\textwidth]{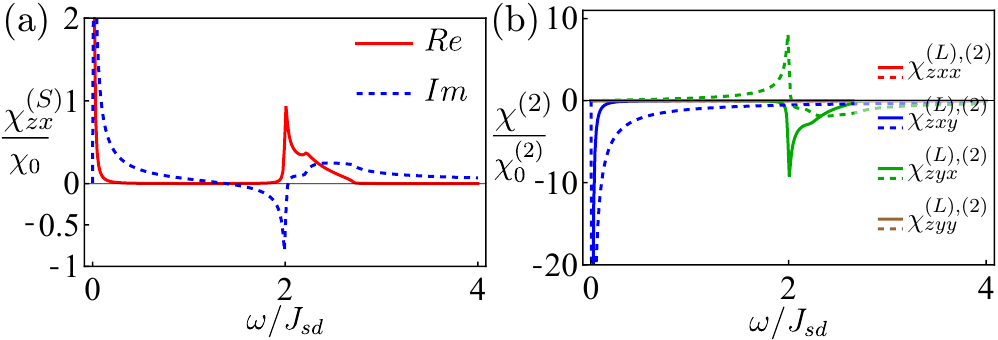}
\caption{
Panels (a) and (b) show the spin Edelstein response coefficient $\chi_{zx}^{(S)}$ and the orbital angular momentum response coefficients $\chi_{ijl}^{(L),(2)}$, respectively, as functions of the driving frequency $\omega$. 
We used the Kubo formulas \eqref{kubo-chiO-1} and \eqref{kubo-chiO-intra} in panel (a) and the density matrix approach in panel (b). Solid and dashed lines correspond to real and imaginary parts, respectively. In all panels, we fix $\alpha =0.5\,J_{sd}/k^{\star}$ with $k^{\star}=\sqrt{2m J_{sd}}$, $\mu=2\,J_{sd}$, and $\tau = 100\,J_{sd}^{-1}$. The normalization coefficients are $\chi_0=e \sqrt{2m/J_{sd}}$ and $\chi_0^{(2)}=e^3/[4m g_L \mu_B J_{sd}^2 (k^{\star})^2]$.
}
\label{fig:kubo-chizx}
\end{figure}

As we show in Fig.~\ref{fig:kubo-chizx}(a), the intraband response has a Drude peak at $\omega=0$, while the interband response is resonantly enhanced near $\omega=2J_{sd}$. The latter is the minimum of the direct interband gap $2B(\mathbf{k})$. Both contributions are proportional to $\alpha_j$, hence the induced spin polarization is controlled by the projection of the electric field along the $p$-wave spin-splitting direction. The resonant enhancement of the spin Edelstein effect in a $p$-wave magnet is one of the key results of our work.

\indent\textcolor{blue}{\em Nonlinear orbital Edelstein response}---
The orbital response differs qualitatively from its spin counterpart. In the model at hand, the diagonal matrix elements of the orbital magnetic moment vanish, while the off-diagonal elements are finite. Because the equilibrium density matrix is diagonal in the band basis, a finite orbital response requires field-induced interband coherence. These matrix elements contain the interband Berry connection, signifying a relation to the quantum geometry; see also the End Matter and SM~\cite{SM}.

We calculate the momentum-resolved orbital Edelstein response tensor perturbatively, 
\begin{eqnarray}
&&L_{i}(\mathbf{k},\Omega) = \chi_{ij}^{(L),(1)}(\Omega, \bk) E_{j}(\Omega) \nonumber\\
&&+\!\!\sum_{\omega_1+\omega_2=\Omega} \chi_{ijl}^{(L),(2)}(\Omega;\omega_1,\omega_2;\mathbf{k}) E_j(\omega_1)E_l(\omega_2)
+ \ldots
\end{eqnarray}
For a monochromatic field, the second-order term contains a rectified response at $\Omega=0$ and a second-harmonic response at $\Omega=\pm2\omega$.

To first order in the electric field and for $\bm{\alpha}\parallel\hat{\mathbf{x}}$, we have
\begin{equation}
\chi_{zx}^{(L), (1)}(\omega, \bk) = \frac{1}{g_L \mu_B}  
\frac{e^2J_{sd}^2\alpha_x^2 k_y}{m\,B^3(\bk) D_{\omega}(\bk)} \Delta n_F,
\label{eq:dm-m-Mz1-before-symmetry}
\end{equation}
where $\Delta n_F = n_F(\epsilon_{+}) -n_F(\epsilon_{-})$. This response oscillates at the driving frequency but is odd in $k_y$. Consequently, the first-order momentum-integrated response tensor vanishes, resulting in zero net orbital angular momentum.

A nonzero momentum-integrated response appears at second order. The complete expressions for $\chi_{zxx}^{(L),(2)}$, $\chi_{zxy}^{(L),(2)}$, and $\chi_{zyx}^{(L),(2)}$ are given in Eqs.~\eqref{eq:dm-m-Mz2-alpha-y-zero-zxx}--\eqref{eq:dm-m-Mz2-alpha-y-zero-zxy} of the End Matter. The component $\chi_{zxx}^{(L),(2)} \propto k_x k_y$ is even under full momentum inversion but is odd under the individual momentum mirrors and, therefore, integrates to zero. By contrast, the mixed components $\chi_{zxy}^{(L),(2)}$ and $\chi_{zyx}^{(L),(2)}$ contain terms proportional to $k_y\partial_{k_y}\Delta n_F$, which are momentum-even and allow for a finite rectified orbital polarization.

The rectified response therefore requires a field with components both parallel and transverse to the $p$-wave spin-splitting direction. For linearly polarized light, $E_x=E_0\cos{\theta}$ and $E_y=E_0\sin{\theta}$, the response is proportional to $E_0^2\sin2\theta$ and vanishes along the principal axes. This momentum-even quadratic response agrees with the period-averaged orbital texture obtained from Floquet micromotion and shown in the bottom row of Fig.~\ref{fig:micromotion}.

\indent\textcolor{blue}{\em Summary and concluding remarks}---
We show that periodic driving converts the hidden momentum-odd texture of a $p$-wave magnet into complementary spin and orbital responses. In the spin channel, a momentum-even polarization appears at linear order in the electric field and oscillates at the driving frequency, see the upper row of Fig.~\ref{fig:micromotion} and Eq.~\eqref{ate-dis1-tau-app-Sz-harmonic-omega}. Its average over the period vanishes, but its momentum integral gives a finite ac spin polarization. The Kubo response separates this effect into a low-frequency intraband contribution and an interband contribution resonantly enhanced near the minimum direct magnetic gap $\sim 2J_{sd}$, see Eq.~\eqref{kubo-chiO-inter-app-1}.

The orbital response is qualitatively different. At linear order, the induced orbital texture remains momentum odd and vanishes after integrating over momenta. A net orbital angular momentum appears in the second-order rectification processes when the electric field has components both parallel and transverse to the $p$-wave spin-splitting direction, see the lower row of Fig.~\ref{fig:micromotion}. This momentum-even dc response originates from interband coherence encoded in the off-diagonal orbital-moment and Berry-connection matrix elements, providing a connection to the quantum geometry. It is therefore absent in single-band descriptions based only on shifted spin-polarized Fermi surfaces commonly used in the literature.

The resonant frequency dependence of the ac spin polarization near $\omega =2J_{sd}$ and the $\sin{2\theta}$ angular dependence of the rectified orbital response provide distinct experimental signatures of the two mechanisms. These responses suggest driven $p$-wave magnets as potential sources of terahertz-range ac spin polarization and rectified dc orbital polarization, all in the absence of SOC. More generally, the results demonstrate that periodic driving can transform hidden momentum-odd spin textures into macroscopic observables, thus unmasking $p$-wave magnetism.

\textit{Note added.---}
While this manuscript was being finalized, magnetic Bloch oscillations and nonlinear Edelstein magnetization in odd-parity magnets were reported in Ref.~\cite{Habel-Knolle-MagneticBlochOscillations-2026}. That work addresses strong-field Bloch dynamics and higher-harmonic generation, whereas the present work focuses on the linear ac susceptibility, its interband resonance, and nonlinear orbital rectification.
Subsequent to the posting of the present work, optical magnetic switching in odd-parity magnets with SOC was reported in Ref.~\cite{Han-Li-OpticalMagneticSwitching-2026}. That study considers a Floquet realization of a finite period-averaged magnetization under elliptically polarized light, whereas the spin polarization identified here appears in the subperiod dynamics, is linear in the driving field, and does not require SOC.

\begin{acknowledgments}
\indent{\em Acknowledgments}---
P.S. thanks Noejung Park, Carsten Ullrich, and Mohsen Yarmohammadi for useful discussions. J.L. was supported by the Research Council of Norway through Grant No. 353894 and its Centres of Excellence funding scheme Grant No. 262633 “QuSpin.”
\end{acknowledgments}

\bibliography{Library-short}

\clearpage
\begin{appendix}

\setcounter{equation}{0} 
\renewcommand{\theequation}{A\arabic{equation}}

\begin{center}
\textbf{End Matter}
\end{center}

\indent\textcolor{blue}{\em Floquet micromotion.}---
For a periodically driven Hamiltonian, we define the following evolution operator:
\begin{equation}
U_{\mathbf{k}}(t,0) = \mathcal{T}_o\exp\left[ -i\int_0^t H_{\mathbf{k}}(t')\,dt' \right],
\label{eq:U_def}
\end{equation}
where $\mathcal{T}_o$ is the time ordering. The Floquet modes at $t=0$ and their quasienergies follow from
\begin{equation}
U_{\mathbf{k}}(T_d,0)|\phi_n(\mathbf{k},0)\rangle = e^{-i\varepsilon_n(\mathbf{k})T_d} |\phi_n(\mathbf{k},0)\rangle.
\label{eq:floquet_eig}
\end{equation}
Their intraperiod evolution is defined by
\begin{equation}
|\phi_n(\mathbf{k},t)\rangle = e^{i\varepsilon_n(\mathbf{k})t} U_{\mathbf{k}}(t,0)|\phi_n(\mathbf{k},0)\rangle,
\label{eq:floquet-mode-evolution}
\end{equation}
which satisfies
$|\phi_n(\mathbf{k},t+T_d)\rangle =|\phi_n(\mathbf{k},t)\rangle$.

For an operator $\hat O(\mathbf{k},t)$, the instantaneous Floquet expectation value is
\begin{equation}
O_n(\mathbf{k},t) = \langle\phi_n(\mathbf{k},t)| \hat O(\mathbf{k},t) |\phi_n(\mathbf{k},t)\rangle.
\label{eq:floquet-observable}
\end{equation}
We define its momentum-even, momentum-odd, and period-averaged
components as
\begin{eqnarray}
\label{eq:parity-and-average}
O_n^{\mathrm{even/odd}}(\mathbf{k},t) &=& \frac{O_n(\mathbf{k},t)
\pm O_n(-\mathbf{k},t)}{2},\\
\label{eq:Sz_timeavg}
\langle\langle \hat{O}\rangle\rangle_n(\mathbf{k}) &=& \frac{1}{T_d}\int_0^{T_d} \langle \hat{O}\rangle_n(\mathbf{k},t)\,dt.
\end{eqnarray}
These definitions are used to obtain the spin textures shown in Fig.~\ref{fig:micromotion}. 

For the orbital texture, we first construct the orbital-moment matrix $\mathbf m^{\mathrm{inst}}(\mathbf{k},t)$ in the instantaneous eigenstate basis of $H_{\mathbf{k}}(t)$. The corresponding operator in the original spin-sector basis reads
\begin{equation}
\hat{\mathbf m}(\mathbf{k},t) = W(\mathbf{k},t) \mathbf{m}^{\mathrm{inst}}(\mathbf{k},t) W^{\dagger}(\mathbf{k},t),
\label{eq:orbital-operator-original-basis}
\end{equation}
where $W(\mathbf{k},t)$ is constructed out of instantaneous eigenvectors. The orbital angular momentum follows from $\mathbf L_n=-\langle\phi_n|\hat{\mathbf m}|\phi_n\rangle/ (g_L\mu_B)$. 

The matrix elements of the orbital magnetization $\mathbf{m}$ contain both diagonal (Abelian) and off-diagonal (non-Abelian) parts~\cite{Cysne-Rappoport-DescriptionOrbitalHall-2025}
\begin{eqnarray}
\label{kubo-mnn1-def}
\mathbf{m}_{nn'} &=& \mathbf{m}_{nn'}^{(0)} + \mathbf{m}_{nn'}^{(1)} +\mathbf{m}_{nn'}^{(2)},\\
\label{kubo-mnn1-0-def}
\mathbf{m}_{nn'}^{(0)} &=& -\frac{ie}{2} \left\langle \partial_{\bk} n \right| \times \left[H - \frac{1}{2}(\epsilon_{n} +\epsilon_{n'})\right] \left| \partial_{\bk} n'\right\rangle,\\
\label{kubo-mnn1-1-def}
\mathbf{m}_{nn'}^{(1)} &=& \frac{e}{4} \left[\left(\bm{\mathcal{A}}_{nn} +\bm{\mathcal{A}}_{n'n'}\right) \times \mathbf{v}_{nn'} \right] (1-\delta_{nn'}),\\
\label{kubo-mnn1-2-def}
\mathbf{m}_{nn'}^{(2)} &=& \frac{e}{4} \left[\left(\mathbf{v}_{nn} +\mathbf{v}_{n'n'}\right) \times \bm{\mathcal{A}}_{nn'} \right] (1-\delta_{nn'}).
\end{eqnarray}
The interband Berry connection $\bm{\mathcal{A}}_{nn'}$ in Eq.~\eqref{kubo-mnn1-2-def} is defined as 
\begin{equation}
\bm{\mathcal{A}}_{nn'}(\mathbf{k},t) = i\bra{u_{n\bk}(t)} \partial_{\bk}\ket{u_{n'\bk}(t)}
\label{eq:A-from-v-omm}
\end{equation}
The orbital magnetic moment given in Eqs.~\eqref{kubo-mnn1-def}--\eqref{kubo-mnn1-2-def} is written as a matrix in instantaneous band space. By transferring to the spin-sectoral basis, we obtain the time-dependent orbital magnetic moment operator determined by the Floquet modes as
\begin{equation}
\langle \hat{\mathbf{m}}\rangle_n(\mathbf{k},t) = \bra{\phi_{n\mathbf{k}}(t)} \hat{\mathbf m}(\mathbf{k},t) \ket{\phi_{n\mathbf{k}}(t)}.
\label{eq:mu-a-omm}
\end{equation}

\indent\textcolor{blue}{\em Kubo linear response approach.}---
The spin and orbital angular momentum response is defined as $O_i = \chi_{ij}^{(O)} E_j$ with $\mathbf{O}$ being spin $\bm{S}$ or angular momentum $\bm{L}$ polarizations. The response tensor is obtained from the Kubo linear response approach
\begin{eqnarray}
\label{kubo-chiO-def}
&&\chi_{ij}^{(O)}(i\Omega_m) \nonumber\\
&&= -\frac{T}{i\Omega_m} \sum_{i\omega_n} \int \frac{d^2k}{(2\pi)^2} \mbox{Tr}{\left\{ \hat{O}_i G(\bk, i \omega_n +i\Omega_m)  \hat{J}_j G(\bk, i \omega_n)\right\}} \nonumber\\
&&= -\frac{1}{i\Omega_m} \sum_{ab}\int \frac{d^2k}{(2\pi)^2} \frac{n_F(\epsilon_b) -n_F(\epsilon_a)}{i\Omega_m +\epsilon_a -\epsilon_b} O_{i;ab} J_{j;ba},
\end{eqnarray}
where we used the band space with the Green's function in Matsubara space defined as $G_a(\bk, i \omega_n) = 1/\left(i\omega_n -\epsilon_{a}\right)$ and performed the Matsubara summation in the last line. The electric current operator is $\hat{\mathbf{J}} = -e\hat{\mathbf{v}}$ with $\hat{\mathbf{v}} = \partial_{\bk}H$ being the velocity operator. Inter- and intraband parts of Eq.~\eqref{kubo-chiO-def} are given in Eqs.~\eqref{kubo-chiO-1} and \eqref{kubo-chiO-intra}.

\indent\textcolor{blue}{\em Density matrix formalism.}---
The equation of motion (quantum Liouville equation) for the density matrix $\rho_{\mathbf{k}}$ is~\cite{Culcer-MacDonald-InterbandCoherenceResponse-2017}
\begin{equation}
\label{dm-0-eq-def}
\frac{\partial \rho_{\mathbf{k}}}{\partial t} - e\mathbf{E}(t)\cdot\partial_{\bk}\rho_{\mathbf{k}} +i \left[ H_0(\mathbf{k}), \rho_{\mathbf{k}}\right] = I_{\rm coll}[\rho_{\mathbf{k}}],
\end{equation}
where $I_{\rm coll}[\rho_{\mathbf{k}}]$ is the collision integral and we assumed the length gauge. The density matrix can be decomposed as
\begin{equation}
\rho_{\mathbf{k}} = \frac{1}{2}\left(n_{\mathbf{k}}\sigma_0 + \mathbf{s}_{\mathbf{k}}\cdot\boldsymbol{\sigma} \right),
\end{equation}

For the generic two-band model
\begin{equation}
\label{dm-0-H-def}
H_0(\mathbf{k}) = d_0(\mathbf{k})\sigma_0 + \mathbf{d}(\mathbf{k})\cdot\boldsymbol{\sigma}
\end{equation}
and the relaxation time approximation, see the right-hand side in Eq.~\eqref{ate-dis1-S-eff-1}, we obtain
\begin{align}
\label{eq:dm-0-n-eom-generic}
&\frac{\partial n_{\mathbf{k}}}{\partial t} - e\mathbf{E}(t)\cdot\partial_{\bk}n_{\mathbf{k}} = -\frac{n_{\mathbf{k}} -n^{(0)}_{\mathbf{k}}}{\tau},\\
\label{eq:dm-0-s-eom-generic}
&\frac{\partial \mathbf{s}_{\mathbf{k}}}{\partial t} -e\mathbf{E}(t)\cdot\partial_{\bk}\mathbf{s}_{\mathbf{k}} = 2\,\mathbf{d}(\mathbf{k})\times\mathbf{s}_{\mathbf{k}} -\frac{\mathbf{s}_{\mathbf{k}} -\mathbf{s}^{(0)}_{\mathbf{k}}}{\tau}.
\end{align}
Here, the equilibrium density matrix is
\begin{equation}
n^{(0)}_{\mathbf{k}} = n_F(\epsilon_+) + n_F(\epsilon_-), \quad \mathbf{s}^{(0)}_{\mathbf{k}} = \frac{\mathbf{d}(\mathbf{k})}{d(\bk)} \Delta n_F.
\label{eq:dm-0-rho-n-s}
\end{equation}

The $n$th-order momentum-integrated orbital angular momentum is
\begin{equation}
L_z^{(n)}(t) = -\frac{1}{g_L \mu_B} \sum_{\eta=\pm} \int \frac{d^2k}{(2\pi)^2} \, \mbox{Tr}\left\{\hat m^{\rm band}_{z}(\mathbf{k})\rho^{(n), {\rm band}}_{\mathbf{k}}(t) \right\},
\end{equation}
where $\rho^{\rm band}_{\mathbf{k}} = W^{\dagger} \rho_{\mathbf{k}} W$ with $W$ being the matrix that diagonalizes the unperturbed Hamiltonian.

We treat the electric field as a perturbation and solve Eqs.~\eqref{eq:dm-0-n-eom-generic} and \eqref{eq:dm-0-s-eom-generic} iteratively. The explicit first- and second-order solutions and the nonlinear response tensors are given in the SM~\cite{SM}. The resulting second-order rectified response tensors are
\begin{align}
\label{eq:dm-m-Mz2-alpha-y-zero-zxx}
&\chi_{zxx}^{L, (2)}(0; \omega,-\omega; \bk) \nonumber\\
&= -\frac{1}{g_L \mu_B}
\frac{e^3J_{sd}\alpha_x k_y}{2m\, B^2(\bk) D_0(\bk)} 
\Bigg\{2 J_{sd} \alpha_x^3 k_x \nonumber\\
&\times \left[ \frac{ \gamma_0 +2\gamma_{-\omega} }{B^3(\bk) D_{-\omega}(\bk) } + \frac{8 \left( \gamma_0+\gamma_{-\omega} \right) }{B(\bk) D_{-\omega}^2(\bk) } \right] \Delta n_F \nonumber\\ 
&- \frac{ 2J_{sd}\alpha_x }{B(\bk)} \left[ \frac{ \gamma_0+\gamma_{-\omega} }{D_{-\omega}(\bk) } + \frac{1}{\gamma_{-\omega}} \right] \partial_{k_x} \Delta n_F \Bigg\},
\end{align}
\begin{align}
\label{eq:dm-m-Mz2-alpha-y-zero-zyx}
&\chi_{zyx}^{L, (2)}(0; \omega,-\omega; \bk) \nonumber\\
&= \frac{1}{g_L \mu_B} \frac{e^3J_{sd}^2\alpha_x^2 k_y}{m\,B^3(\bk) D_{0}(\bk)} 
\frac{ \gamma_{0} + \gamma_{-\omega}}{D_{-\omega}(\bk)} 
\partial_{k_y} \Delta n_F,\\
\label{eq:dm-m-Mz2-alpha-y-zero-zxy}
&\chi_{zxy}^{L, (2)}(0; \omega,-\omega; \bk) \nonumber\\
&= \frac{1}{g_L \mu_B} \frac{e^3J_{sd}^2\alpha_x^2 k_y}{m\,B^3(\bk) D_{0}(\bk)} 
\frac{1}{\gamma_{-\omega}}
\partial_{k_y} \Delta n_F.
\end{align}

\end{appendix}

\end{document}



\title{Supplemental Material for \textit{Dynamical unmasking of $p$-wave magnetism by ac spin and dc orbital polarizations}
}

\author{Yantao Li}
\affiliation{Department of Physics and Astronomy, University of Missouri, Columbia, Missouri, 65211, USA}

\author{Jacob Linder}
\affiliation{Center for Quantum Spintronics, Department of Physics, Norwegian University of Science and Technology, NO-7491 Trondheim, Norway}

\author{Pavlo~Sukhachov}
\email{pavlo.sukhachov@missouri.edu}
\affiliation{Department of Physics and Astronomy, University of Missouri, Columbia, Missouri, 65211, USA}
\affiliation{MU Materials Science \& Engineering Institute,
University of Missouri, Columbia, Missouri, 65211, USA}

\maketitle
\tableofcontents

\section{Models}
\label{sec:models}
In this section, we introduce lattice and continuum models of a $p$-wave magnet.

\subsection{Lattice model}
\label{sec:models-lattice}

We use the following minimal lattice model for a $p$-wave magnet~\cite{Brekke-Linder-MinimalModelsTransport-2024}
\begin{eqnarray}
\label{eq:Ht}
H(\bk) &=&  \xi_{\mathbf{k}}\,\sigma_0\otimes\tau_0 + M(\mathbf{k})\,\sigma_z\otimes\tau_0 + J_{sd}\,\sigma_x\otimes\tau_z \nonumber\\
&=& -2\lambda[\cos\!\big(k_x a\big)
+ \cos\!\big(k_y a \big)]
\sigma_0\otimes\tau_0 \nonumber\\
&+& \sum_{i=x,y} \tilde{\alpha}_i\sin{(k_i a)} \sigma_z\otimes\tau_0+ J_{sd}\,\sigma_x\otimes\tau_z.
\end{eqnarray}
Here, $\lambda$ is the nearest-neighbor hopping amplitude, $\tilde{\bm{\alpha}}$ is the spin-splitting strength, and $J_{sd}$ is the $sd$ exchange coupling strength. The Pauli matrices $\bm{\sigma}$ and $\bm{\tau}$ act in the spin and sector spaces~\cite{Brekke-Linder-MinimalModelsTransport-2024}, respectively. 

The eigenfunctions and eigenenergies of the Hamiltonian in Eq.~\eqref{eq:Ht} are given by
\begin{eqnarray}
\label{ate-psi}
\left| n\right \rangle = 
\frac{1}{\sqrt{1 + \frac{J_{sd}^2}{\left[\xi_k -M(\bk) -\epsilon_n\right]^2}}} \begin{pmatrix}
1\\
\frac{\eta J_{sd}}{\epsilon_n + M(\bk) -\xi_k}
\end{pmatrix}
\end{eqnarray}
and
\begin{equation}
\label{ate-eps}
\epsilon_{\bk,\pm} = \xi_k \pm \sqrt{M^2(\bk) +J_{sd}^2},
\end{equation}
respectively. Here, $n=1$ and $n=2$ correspond to the $-$ and $+$ signs, respectively, in Eq.~\eqref{ate-eps} and $\eta=+$. The other sector, $\eta=-$, is denoted by $n=3,4$.

\subsection{Continuum model}
\label{sec:models-continuum}

To make an analytical advance, we also use a continuum model derived by expanding the Hamiltonian \eqref{eq:Ht} around the $\Gamma$ point, absorbing $-4\lambda$ into the definition of the chemical potential, and projecting onto a sector $\eta=\pm$,
\begin{equation}
\label{ate-H-def}
H(\bk) = \xi_{\bk} \sigma_0 +M(\bk) \sigma_z +\eta J_{sd} \sigma_x,
\end{equation}
where $\xi_{\bk}=k^2/(2m)$ is the kinetic energy with $(2m)^{-1}=\lambda a^2$ and $M(\bk)=(\bm{\alpha}\cdot\bk)$ is the spin-splitting term with $a\tilde{\bm{\alpha}}=\bm{\alpha}$. Here and in the rest of the Supplemental Material, we set $\hbar=1$.

The eigenfunctions and eigenenergies of the Hamiltonian in Eq.~\eqref{ate-H-def} are given by Eqs.~\eqref{ate-psi} and \eqref{ate-eps}, respectively, with $\xi_{\bk}=k^2/(2m)$ and $M(\bk)=(\bm{\alpha}\cdot\bk)$.

The equilibrium spin polarization of the bands is
\begin{equation}
\label{ate-S0nn}
\left\langle n\right| \hat{S}_{z}\left| n\right \rangle = \mp \frac{1}{2} \frac{(\bm{\alpha}\cdot\mathbf{k})}{\sqrt{(\bm{\alpha}\cdot\mathbf{k})^2 +J_{sd}^2}},
\end{equation}
where the spin operator is $\hat{\mathbf{S}} = \bm{\sigma}/2$ and the upper (lower) sign is valid for $n=1$ ($n=2$). 

We find it convenient to introduce the following characteristic scales:
\begin{eqnarray}
\label{eq:dm-m-dimless}
k^{\star} = \sqrt{2mJ_{sd}}, \quad \alpha_{0} = \frac{J_{sd}}{k^{\star}}, \quad m_{z,0} = \frac{e}{4m}, \quad \tau_0=J_{sd}^{-1}, \quad E_0 = \frac{J_{sd} k^{\star}}{e}.
\end{eqnarray}

\begin{figure}[t]
   \centering
   \includegraphics[width=6.9in]{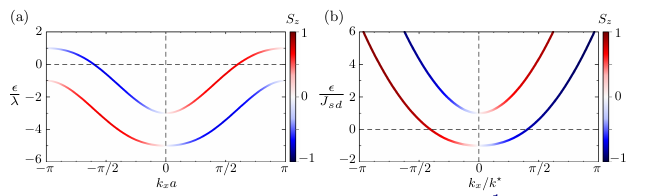}
   \caption{The energy spectra of the lattice [panel (a)] and continuum
   [panel (b)] models at $k_y=0$. The parameters are
   $J_{sd}=\widetilde\alpha_x=\lambda$ and
   $\widetilde\alpha_y=0$ in panel (a), and $\alpha_x=\alpha_0$ and
   $\alpha_y=0$ in panel (b). Both sectors $\eta=\pm1$ are included.}
   \label{fig1_SM}
\end{figure}

The band spectra of the lattice and continuum models are shown in Fig.~\ref{fig1_SM}.

\section{Floquet approach and micromotion}
\label{sec:Floquet}

\subsection{Floquet--Sambe formulation}

For a periodically driven system with $H_{\bk}(t+T_d)=H_{\bk}(t)$, where $T_d=2\pi/\omega$, we apply Floquet theory~\cite{Shirley1965,Zeldovich1967,Sam73a,Oka-Kitamura-FloquetEngineeringQuantum-2019}. The solutions of the time-dependent Schr\"odinger equation can be written as $\ket{\psi_\nu(\bk,t)}=e^{-i\varepsilon_\nu(\bk)t}\ket{\phi_\nu(\bk,t)}$ and $\ket{\phi_\nu(\bk,t+T_d)}=\ket{\phi_\nu(\bk,t)}$, where $\nu$ labels the Floquet bands and $\varepsilon_\nu(\bk)$ denotes the quasienergy. Substitution into the Schr\"odinger equation gives
\begin{equation}
\left[H_{\bk}(t)-i\partial_t\right]
\ket{\phi_\nu(\bk,t)}
=
\varepsilon_\nu(\bk)
\ket{\phi_\nu(\bk,t)}.
\end{equation}
The quasienergies are defined modulo $\omega$, and we choose the first Floquet Brillouin zone $(-\omega/2,\omega/2]$. We use the Fourier convention
\begin{equation}
H_{\bk}(t)
=
\sum_{\ell\in\mathbb{Z}}
H_{\bk}^{(\ell)}e^{-i\ell\omega t},
\qquad
\ket{\phi_\nu(\bk,t)}
=
\sum_{m\in\mathbb{Z}}
e^{-im\omega t}\widetilde{\Phi}_{\nu,m}(\bk).
\label{eq:floquet-fourier-convention_SM}
\end{equation}
In our system, the electromagnetic field is introduced through the Peierls substitution $\bk\longrightarrow\bk+e\mathbf A(t)$,
where $-e<0$ is the electron charge. We take $\mathbf A(t)=\left\{A_x\cos(\omega t),A_y\cos(\omega t+\varphi)\right\}$, where $\varphi$ controls the light polarization.

The Floquet matrix can be written as
\begin{equation}
\mathcal H_{F,\bk}
=
\begin{pmatrix}
\ddots & \vdots & \vdots & \vdots & \iddots\\
\cdots &
H_{\bk}^{(0)}+\omega I_4 &
H_{\bk}^{(-1)} &
H_{\bk}^{(-2)} &
\cdots\\
\cdots &
H_{\bk}^{(1)} &
H_{\bk}^{(0)} &
H_{\bk}^{(-1)} &
\cdots\\
\cdots &
H_{\bk}^{(2)} &
H_{\bk}^{(1)} &
H_{\bk}^{(0)}-\omega I_4 &
\cdots\\
\iddots & \vdots & \vdots & \vdots & \ddots
\end{pmatrix}.
\label{Floquet_matrix_SM}
\end{equation}
Here $I_4$ is the $4\times4$ identity matrix. Equivalently,
\begin{equation}
\left(\mathcal H_{F,\bk}\right)_{mm'}
=
H_{\bk}^{(m-m')}
-m\omega I_4\delta_{mm'}.
\label{eq:floquet-matrix-elements_SM}
\end{equation}

For the lattice model in Eq.~\eqref{eq:Ht}, we have
\begin{align}
H_{\bk}^{(0)}
={}&
-2\lambda
\left[
J_0(eaA_x)\cos(k_xa)
+
J_0(eaA_y)\cos(k_ya)
\right]\sigma_0\otimes\tau_0
\nonumber\\
&+
\left[
\widetilde{\alpha}_xJ_0(eaA_x)\sin(k_xa)
+
\widetilde{\alpha}_yJ_0(eaA_y)\sin(k_ya)
\right]\sigma_z\otimes\tau_0
\nonumber\\
&+
J_{sd}\sigma_x\otimes\tau_z,
\end{align}
where $J_r(x)$ denotes the Bessel function of the first kind. For $\ell\neq0$, the Fourier components can be written as
\begin{align}
H_{\bk}^{(\ell)}
={}&
-2\lambda
\left[
\lambda_x^{(\ell)}(k_xa)
+
\lambda_y^{(\ell)}(k_ya)
\right]\sigma_0\otimes\tau_0
\nonumber\\
&+
\left[
\widetilde{\alpha}_x
\alpha_x^{(\ell)}(k_xa)
+
\widetilde{\alpha}_y
\alpha_y^{(\ell)}(k_ya)
\right]\sigma_z\otimes\tau_0,
\end{align}
where
\begin{align}
\lambda_x^{(\ell)}(k_xa)
={}&
\sum_{r=-\infty}^{\infty}
\left(
\frac{i^re^{ik_xa}}{2}
+
\frac{(-i)^re^{-ik_xa}}{2}
\right)
J_r(eaA_x)\delta_{r,-\ell},
\\
\lambda_y^{(\ell)}(k_ya)
={}&
\sum_{r=-\infty}^{\infty}
\left(
\frac{i^re^{ik_ya}}{2}
+
\frac{(-i)^re^{-ik_ya}}{2}
\right)
J_r(eaA_y)\delta_{r,-\ell}e^{ir\varphi},
\\
\alpha_x^{(\ell)}(k_xa)
={}&
\sum_{r=-\infty}^{\infty}
\left(
\frac{i^{r-1}e^{ik_xa}}{2}
-
\frac{(-i)^{r+1}e^{-ik_xa}}{2}
\right)
J_r(eaA_x)\delta_{r,-\ell},
\\
\alpha_y^{(\ell)}(k_ya)
={}&
\sum_{r=-\infty}^{\infty}
\left(
\frac{i^{r-1}e^{ik_ya}}{2}
-
\frac{(-i)^{r+1}e^{-ik_ya}}{2}
\right)
J_r(eaA_y)\delta_{r,-\ell}e^{ir\varphi}.
\end{align}

This representation makes the role of the photon sectors explicit. The photon-sector index $m\in\mathbb Z$ labels the Fourier harmonic $e^{-im\omega t}$ and adjacent sectors are Floquet replicas separated in quasienergy by $\omega$. The zeroth harmonic $H_{\bk}^{(0)}$ is the time-averaged Hamiltonian, whereas $H_{\bk}^{(\ell\neq0)}$ couples sectors whose indices differ by $\ell$ and is responsible for photon-assisted reconstruction of the quasienergy bands and Floquet wave functions.

\subsection{Floquet micromotion}
\label{sec:micromotion}

In a periodically driven quantum system, micromotion describes the dynamics within each driving period. The time-evolution operator can be written as
\begin{equation}
U_{\bk}(t,0)
=
\mathcal T_o
\exp\left[
-i\int_0^tH_{\bk}(t')\,dt'
\right],
\label{eq:U_def_SM}
\end{equation}
where $\mathcal T_o$ denotes time ordering. 

The Floquet spectrum is obtained by diagonalizing the one-period evolution operator $U_{\bk}(T_d,0)$. The Floquet modes at $t=0$ and their quasienergies satisfy
\begin{equation}
U_{\bk}(T_d,0)
\lvert\phi_\nu(\bk,0)\rangle
=
e^{-i\varepsilon_\nu(\bk)T_d}
\lvert\phi_\nu(\bk,0)\rangle.
\label{eq:floquet_eig_SM}
\end{equation}
Thus, numerical diagonalization of $U_{\bk}(T_d,0)$ provides the quasienergies and Floquet modes at $t=0$. The corresponding physical Floquet states at an arbitrary time are
\begin{equation}
\lvert\psi_\nu(\bk,t)\rangle
=
U_{\bk}(t,0)\lvert\phi_\nu(\bk,0)\rangle
=
e^{-i\varepsilon_\nu(\bk)t}
\lvert\phi_\nu(\bk,t)\rangle,
\end{equation}
where
\begin{equation}
\lvert\phi_\nu(\bk,t)\rangle
=
e^{i\varepsilon_\nu(\bk)t}
U_{\bk}(t,0)
\lvert\phi_\nu(\bk,0)\rangle
\end{equation}
satisfies
\begin{equation}
\lvert\phi_\nu(\bk,t+T_d)\rangle
=
\lvert\phi_\nu(\bk,t)\rangle.
\end{equation}

In what follows, we discuss two methods for investigating the spin polarization of the bands, and present the corresponding results.

\subsubsection{Method 1: time average}
\label{sec:micromotion-1}

This direct time-average approach provides an independent benchmark for the truncated Floquet-matrix calculation. We divide one driving period into $N_t$ intervals, with
\begin{equation}
\Delta t=\frac{T_d}{N_t},
\qquad
t_j=j\Delta t,
\qquad
j=0,\ldots,N_t-1.
\end{equation}
The short-time propagator over the interval $[t_j,t_{j+1}]$ is approximated by
\begin{equation}
U_{\bk,j}
\equiv
U_{\bk}(t_{j+1},t_j)
\simeq
\exp\left[
-iH_{\bk}\left(t_j+\frac{\Delta t}{2}\right)\Delta t
\right],
\end{equation}
where the Hamiltonian is evaluated at the midpoint of each interval. The one-period evolution operator is then approximated by
\begin{equation}
U_{\bk}(T_d,0)
\approx
U_{\bk,N_t-1}\cdots
U_{\bk,1}U_{\bk,0}.
\end{equation}

The intraperiod dynamics of each Floquet state can be reconstructed using the same short-time propagators. Taking the Floquet mode at $t=0$ as the initial state,
\begin{equation}
\lvert\psi_\nu(\bk,0)\rangle
=
\lvert\phi_\nu(\bk,0)\rangle,
\end{equation}
each state is propagated recursively according to
\begin{equation}
\lvert\psi_\nu(\bk,t_{j+1})\rangle
=
U_{\bk,j}
\lvert\psi_\nu(\bk,t_j)\rangle.
\end{equation}
Equivalently,
\begin{equation}
\lvert\psi_\nu(\bk,t_j)\rangle
=
U_{\bk,j-1}\cdots U_{\bk,1}U_{\bk,0}
\lvert\phi_\nu(\bk,0)\rangle.
\label{eq:micromotion_SM}
\end{equation}

We define the instantaneous spin polarization as
\begin{equation}
\langle S_z\rangle_\nu(\bk,t)
=
\langle\phi_\nu(\bk,t)|
\hat S_z
|\phi_\nu(\bk,t)\rangle
=
\langle\psi_\nu(\bk,t)|
\hat S_z
|\psi_\nu(\bk,t)\rangle.
\label{eq:Sz_inst_SM}
\end{equation}

Since the Floquet state $|\psi_\nu(\bk,t)\rangle$ and the periodic Floquet mode $|\phi_\nu(\bk,t)\rangle$ differ only by a global phase, they yield the same expectation value. The period-averaged spin polarization is
\begin{equation}
\langle\!\langle S_z\rangle\!\rangle_\nu(\bk)
=
\frac{1}{T_d}
\int_0^{T_d}
\langle S_z\rangle_\nu(\bk,t)\,dt.
\label{eq:Sz_timeavg_SM}
\end{equation}
Numerically,
\begin{equation}
\langle\!\langle S_z\rangle\!\rangle_\nu(\bk)
\simeq
\frac{1}{N_t}
\sum_{j=0}^{N_t-1}
\langle S_z\rangle_\nu(\bk,t_j).
\end{equation}

\subsubsection{Method 2: summation over the Fourier components}
\label{sec:micromotion-2}
The period-averaged spin polarization can be obtained from the Fourier components of the Floquet modes. We truncate the Sambe space at $|m|\leq N_F$, so that $2N_F+1$ Fourier sectors are retained:
\begin{equation}
\begin{split}
\langle\!\langle S_z\rangle\!\rangle_\nu(\bk)
=
\sum_{m=-\infty}^{\infty}
\widetilde\Phi_{\nu,m}^{\dagger}(\bk)
\hat S_z
\widetilde\Phi_{\nu,m}(\bk) \approx
\sum_{m=-N_F}^{N_F}
\widetilde\Phi_{\nu,m}^{\dagger}(\bk)
\hat S_z
\widetilde\Phi_{\nu,m}(\bk).
\end{split}
\label{eq:Sz_fourier_identity_SM}
\end{equation}
This provides an equivalent frequency-domain evaluation of the period-averaged spin polarization.

Here $\widetilde\Phi_{\nu,m}(\bk)$ denotes the component of the Floquet eigenvector labeled by $\nu$ in the $m$th Fourier sector, as defined in Eq.~\eqref{eq:floquet-fourier-convention_SM}. The full Floquet eigenvector is therefore written as
\begin{equation}
\widetilde\Phi_\nu(\bk)
=
\begin{pmatrix}
\vdots\\
\widetilde\Phi_{\nu,-1}(\bk)\\
\widetilde\Phi_{\nu,0}(\bk)\\
\widetilde\Phi_{\nu,1}(\bk)\\
\vdots
\end{pmatrix}.
\end{equation}
For the original $4\times4$ lattice Hamiltonian, each $\widetilde\Phi_{\nu,m}(\bk)$ is a four-component spinor in spin and sectoral space. The summation over $m$ in Eq.~\eqref{eq:Sz_fourier_identity_SM} therefore adds the contribution from every retained photon sector.

\subsubsection{Floquet gauge}
\label{sec:micromotion-3}

The choice of the time origin fixes the phase of the periodic drive and therefore defines the Floquet gauge. For two reference times $t_0$ and $t_0'$, define
\begin{equation}
U_{F,\bk}(t_0)
=
U_{\bk}(t_0+T_d,t_0).
\end{equation}
The corresponding one-period operators are related by
\begin{equation}
U_{F,\bk}(t_0')
=
U_{\bk}(t_0',t_0)
U_{F,\bk}(t_0)
U_{\bk}^{\dagger}(t_0',t_0).
\label{eq:floquet_gauge_SM}
\end{equation}
Therefore, the quasienergy spectrum is independent of the time origin, whereas the effective Floquet Hamiltonian, Floquet eigenvectors, and micromotion depend on the chosen Floquet gauge. Changing the time origin shifts the phase-resolved intraperiod dynamics but leaves observables averaged over a complete driving period unchanged. Throughout this Supplemental Material, we fix the time origin at $t_0=0$.

\subsubsection{Results for time evolution of spin polarization}
\label{sec:micromotion-4}

\begin{figure}[t]
   \centering
   \includegraphics[width=6.7in]{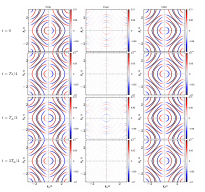}
   \caption{
    Floquet micromotion of the spin polarization at $t=0$, $t=0.25\,T_d$, $t=0.5\,T_d$, and $t=0.75\,T_d$, shown from top to bottom. The parameters are $eA_x=eA_y=0.1\,a^{-1}$, $J_{sd}=\tilde{\alpha}_x=\lambda$, $\tilde{\alpha}_y=0$, $\omega = 2\pi J_{sd}/4$, and $\varphi=0$.}
   \label{fig2_SM}
\end{figure}

\begin{figure}[t]
   \centering
   \includegraphics[width=6.7in]{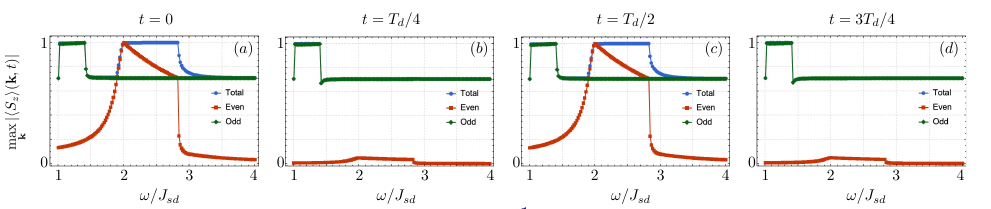}
   \caption{
    Frequency dependence of the spin polarization during Floquet micromotion. The parameters are set to $eA_x=eA_y=0.1\,a^{-1}$, $J_{sd}=\tilde{\alpha}_x=\lambda$, $\tilde{\alpha}_y=0$, and $\varphi=0$.
    }
   \label{fig3_SM}
\end{figure}

For the lattice model, we identify the parity of the driven spin texture by comparing the spin polarization at $\mathbf{k}$ and $-\mathbf{k}$. The even and odd components are defined as follows:
\begin{equation}
  \langle S_z^{\mathrm{even}}\rangle_n(\mathbf{k},t)  =\frac{\langle S_z\rangle_n(\mathbf{k},t)+\langle S_z\rangle_n(-\mathbf{k},t)}{2},
\end{equation}
and
\begin{equation}
  \langle S_z^{\mathrm{odd}}\rangle_n(\mathbf{k},t)  =\frac{\langle S_z\rangle_n(\mathbf{k},t)-\langle S_z\rangle_n(-\mathbf{k},t)}{2}.
\end{equation}
Note that we denote the sum of the even and odd components as the total spin expectation value, $\langle S_z^{\mathrm{total}}\rangle_n(\mathbf{k},t)\equiv\langle S_z\rangle_n(\mathbf{k},t)$. We show the micromotion of spin polarization in Fig.~\ref{fig2_SM}. Alongside the formation of Floquet sidebands, the spin polarization acquires an even-in-momentum component during the intraperiod evolution. Both even- and odd-in-momentum components vary periodically in response to the drive. Consequently, the nodal line near the $\Gamma$ point undergoes periodic motion over the driving cycle. To visualize the frequency dependence of the spin polorizatioon during the micromotion, we define the maximum value of the spin polorization in the momentum space as $\max_{\mathbf{k}} |\langle S_z\rangle(\mathbf{k},t)|$. The summed over two sectors odd-in-momentum component maintains a maximum value near $0.7$ across frequencies and during the micromotion. In contrast, the even-in-momentum component exhibits a distinct resonance peak around $\omega= 2J_{sd}$, outside of which it is suppressed, see Fig.~\ref{fig3_SM}. This resonant amplitude depends on time $t$, attaining the maximal value at $t = 0$ and $0.5T_d$, while reaching the minimal value at $t = 0.25T_d$ and $0.75T_d$.
 
\subsection{Orbital magnetic and angular moments}
\label{sec:om-Floquet}

\subsubsection{Definitions}
\label{sec:om-def}

The orbital magnetic moment used in the micromotion calculation is first constructed in the instantaneous band basis. At fixed $\bk$ and $t$, the instantaneous eigenstates satisfy
\begin{equation}
H_{\bk}(t)|u_{a\bk}(t)\rangle
=
\epsilon_{a\bk}(t)|u_{a\bk}(t)\rangle.
\label{eq:inst-eigen-omm_SM}
\end{equation}

The velocity operators are
\begin{equation}
\hat v_i(\bk,t)
=
\partial_{k_i}H_{\bk}(t),
\qquad
i=x,y.
\end{equation}

For the lattice Hamiltonian in Eq.~\eqref{eq:Ht}, we define
\begin{equation}
\kappa_i(\bk,t)
=
a[k_i+eA_i(t)].
\end{equation}
The velocity components are
\begin{align}
\hat v_x(\bk,t)
={}&
a\left[
2\lambda\sin\kappa_x(\bk,t)\,
\sigma_0\otimes\tau_0
+
\widetilde\alpha_x\cos\kappa_x(\bk,t)\,
\sigma_z\otimes\tau_0
\right],
\\
\hat v_y(\bk,t)
={}&
a\left[
2\lambda\sin\kappa_y(\bk,t)\,
\sigma_0\otimes\tau_0
+
\widetilde\alpha_y\cos\kappa_y(\bk,t)\,
\sigma_z\otimes\tau_0
\right].
\end{align}
The corresponding matrix elements are
\begin{equation}
v_{ab}^{i}(\bk,t)
=
\langle u_{a\bk}(t)|
\hat v_i(\bk,t)
|u_{b\bk}(t)\rangle.
\label{eq:v-matrix-omm_SM}
\end{equation}

The Berry-connection matrix is
\begin{equation}
\bm{\mathcal A}_{ab}(\bk,t)
=
i\langle u_{a\bk}(t)|
\partial_{\bk}u_{b\bk}(t)\rangle.
\end{equation}
For $a\neq b$, its interband components can be evaluated as
\begin{equation}
\bm{\mathcal A}_{ab}(\bk,t)
=
\frac{\mathbf v_{ab}(\bk,t)}
{i[\epsilon_{a\bk}(t)-\epsilon_{b\bk}(t)]}.
\label{eq:A-from-v-omm_SM}
\end{equation}

Following Ref.~\cite{Cysne-Rappoport-DescriptionOrbitalHall-2025},
we set $\hbar=1$ while keeping $e$ explicit and write
\begin{equation}
m_{ab}^{z}(\bk,t)
=
m_{ab}^{(0),z}(\bk,t)
+
m_{ab}^{(1),z}(\bk,t)
+
m_{ab}^{(2),z}(\bk,t).
\label{eq:m-total-omm_SM}
\end{equation}
Since
\begin{equation}
|\partial_{k_i}u_{b\bk}(t)\rangle
=
\sum_c
|u_{c\bk}(t)\rangle
\langle u_{c\bk}(t)|
\partial_{k_i}u_{b\bk}(t)\rangle,
\end{equation}
the zero-order term can be written as
\begin{align}
m_{ab}^{(0),z}(\bk,t)
={}&
-\frac{ie}{2}
\sum_c
\left[
\epsilon_{c\bk}(t)
-
\frac{
\epsilon_{a\bk}(t)+\epsilon_{b\bk}(t)
}{2}
\right]
\nonumber\\
&\times
\left[
\mathcal A_{ac}^{x}(\bk,t)
\mathcal A_{cb}^{y}(\bk,t)
-
\mathcal A_{ac}^{y}(\bk,t)
\mathcal A_{cb}^{x}(\bk,t)
\right].
\label{eq:m0-omm_SM}
\end{align}
The first-order term is
\begin{align}
m_{ab}^{(1),z}(\bk,t)
={}&
\frac{e}{4}
\left\{
[\mathcal A_{aa}^{x}(\bk,t)+\mathcal A_{bb}^{x}(\bk,t)]
v_{ab}^{y}(\bk,t)
\right.
\nonumber\\
&\left.
-
[\mathcal A_{aa}^{y}(\bk,t)+\mathcal A_{bb}^{y}(\bk,t)]
v_{ab}^{x}(\bk,t)
\right\}
(1-\delta_{ab}).
\label{eq:m1-omm_SM}
\end{align}
The second-order term is
\begin{align}
m_{ab}^{(2),z}(\bk,t)
={}&
\frac{e}{4}
\left\{
[v_{aa}^{x}(\bk,t)+v_{bb}^{x}(\bk,t)]
\mathcal A_{ab}^{y}(\bk,t)
\right.
\nonumber\\
&\left.
-
[v_{aa}^{y}(\bk,t)+v_{bb}^{y}(\bk,t)]
\mathcal A_{ab}^{x}(\bk,t)
\right\}
(1-\delta_{ab}).
\label{eq:m2-omm_SM}
\end{align}

Thus,
\begin{equation}
m_z^{\mathrm{inst}}(\mathbf{k},t) =
\begin{pmatrix}
m^z_{11} & m^z_{12} & \cdots & m^z_{1N}\\
m^z_{21} & m^z_{22} & \cdots & m^z_{2N}\\
\vdots & \vdots & \ddots & \vdots\\
m^z_{N1} & m^z_{N2} & \cdots & m^z_{NN}
\end{pmatrix}_{\mathrm{inst}}.
\label{eq:m-inst-matrix-omm_SM}
\end{equation}
Here, $N=4$, and $m_z^{\mathrm{inst}}(\mathbf{k},t)$
represents the orbital magnetic-moment operator in the
instantaneous band basis.
To evaluate its expectation value in the time-evolving
Floquet states expressed in the original spin-sector basis,
we first transform the operator to that basis.
For this purpose, we introduce the unitary matrix
\begin{equation}
W(\bk,t)
=
\left(
|u_{1\bk}(t)\rangle,
|u_{2\bk}(t)\rangle,
|u_{3\bk}(t)\rangle,
|u_{4\bk}(t)\rangle
\right),
\end{equation}
whose columns are the instantaneous eigenvectors expressed in the original spin-sector basis. The orbital magnetic-moment operator in this basis is
\begin{equation}
\hat m_z(\bk,t)
=
W(\bk,t)
m_z^{\mathrm{inst}}(\bk,t)
W^\dagger(\bk,t).
\label{eq:m-original-basis-omm_SM}
\end{equation}

The orbital angular momentum is defined via the relation $\mathbf{L} = -(1/g_L \mu_B) \mathbf{m}$, where $g_L=1$ is the Land\'{e} $g$-factor and $\mu_B = e/(2m_e)$ is the Bohr magneton and $\mathbf{m}$ is the orbital magnetic moment. Its momentum-even and momentum-odd components in $z$ direction are
\begin{equation}
L_{z,\nu}^{\mathrm{even/odd}}(\bk,t)
=
\frac{
L_{z,\nu}(\bk,t)
\pm
L_{z,\nu}(-\bk,t)
}{2}.
\label{eq:Lz-parity_SM}
\end{equation}

\subsubsection{Results for the time evolution of orbital angular momentum}
\label{sec:om-results}

\begin{figure}[t]
   \centering
   \includegraphics[width=6.7in]{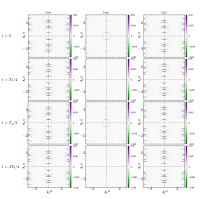}
   \caption{The Floquet micromotion of orbital angular momentum at $t=0$, $t=0.25\,T_d$, $t=0.5\,T_d$, and $t=0.75\,T_d$, shown from top to bottom. The parameters are set to $eA_x=eA_y=0.1\,a^{-1}$, $J_{sd}=\tilde{\alpha}_x=\lambda$, $\tilde{\alpha}_y=0$, $\omega = 2\pi J_{sd}/4$, and $\varphi=0$. The results for the even part have been multiplied by a factor of $3$.
   }
   \label{fig4_SM}
\end{figure}

We show the micromotion of orbital angular momentum for the lattice model in Fig.~\ref{fig4_SM}. We used Eqs.~\eqref{eq:m-total-omm_SM}--\eqref{eq:m-original-basis-omm_SM} and Eq.~\eqref{eq:Lz-parity_SM}. The leading momentum-odd component oscillates at $\omega$. The momentum-even component appears at second order and contains a rectified dc part and a $2\omega$ harmonic. In contrast to the micromotion of the spin polarization shown in Fig.~\ref{fig2_SM}, no pronounced oscillating nodal line is observed because the generation of the even component of the orbital angular momentum is a second-order effect and is, therefore, relatively small. After averaging over one driving period, the orbital-angular-momentum texture is exactly even, in sharp contrast to the spin-polarization texture, which is odd under the same averaging procedure. Note that the orbital magnetic-moment operator is constructed from Eqs.~\eqref{eq:m-total-omm_SM}--\eqref{eq:m-original-basis-omm_SM}. The orbital angular momentum and its even and odd components are evaluated using \eqref{eq:Lz-parity_SM}.

\begin{figure}[t]
   \centering
   \includegraphics[width=6.9in]{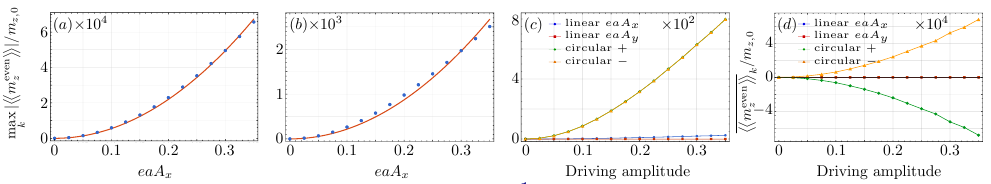}
   \caption{
    Panels (a) and (b) show the maximum absolute value of the orbital magnetic moment averaged over one driving period as a function of $eaA_x$. The parameters are $eA_y=0.1\,a^{-1}$, $J_{sd}=\tilde{\alpha}_x=\lambda$, $\tilde{\alpha}_y=0$, and $\varphi=0$. The driving frequencies are (a) $\omega=2\pi J_{sd}$ and (b) $\omega= 2\pi J_{sd}/4$. The red lines show quadratic fits proportional to $(eaA_x)^2$. Panel (c) shows the maximal absolute value of the orbital magnetic moment averaged over one period, and panel (d) shows the net orbital magnetic moment averaged over both momentum and one period. Both are plotted as functions of the driving amplitude for $x$- and $y$-linear polarizations ($eaA_x$ and $eaA_y$, respectively) and for circular polarizations labeled $+$ ($\varphi=\pi/2$) and $-$ ($\varphi=-\pi/2$). In panels (c) and (d), $\omega= 2\pi J_{sd}/4$. The values in panels (a)--(d) are multiplied by $10^4$, $10^3$, $10^2$, and $10^4$, respectively.}
   \label{figME_SM}
\end{figure}

Figure~\ref{figME_SM} shows the maximum absolute value of the orbital magnetic moment averaged over one period as a function of $eaA_x$. The quadratic scaling with the amplitude indicates that the even part of the orbital magnetic moment arises from a nonlinear effect.

\subsection{Convergence}
\label{sec:micromotion-5}

We assess convergence on a uniform $2000\times2000$ mesh in momentum space covering the first Brillouin zone. For the direct time-domain calculation, results obtained with $N_t=16$ and $32$ are compared with the reference calculation at $N_t^{\mathrm{ref}}=64$. For the Floquet-matrix calculation, we retain the sectors $m=-N_F,\ldots,N_F$ and vary the cutoff over $N_F=2,\ldots,10$, using $N_F^{\mathrm{ref}}=10$ as the reference.

The corresponding Floquet-matrix dimension is $4(2N_F+1)$, which equals $84$ for $N_F=10$.

Because quasienergies are defined modulo $\omega$, we quantify their convergence using
\begin{equation}
\delta_\varepsilon
=
\max_{\bk,\nu}
\frac{1}{\lambda}
\min_{q\in\mathbb Z}
\left|
\varepsilon_\nu^{(1)}(\bk)
-
\varepsilon_\nu^{(2)}(\bk)
+
q\omega
\right|,
\label{eq:quasienergy-error_SM}
\end{equation}
after matching Floquet bands by maximizing the overlap of their eigenvectors. The spin-polarization error is
\begin{equation}
\delta_{S_z}
=
\max_{\bk,\nu}
\left|
\langle\!\langle S_z\rangle\!\rangle_\nu^{(1)}(\bk)
-
\langle\!\langle S_z\rangle\!\rangle_\nu^{(2)}(\bk)
\right|.
\label{eq:spin-error_SM}
\end{equation}

At $N_t=16$ and $N_F=10$, the maximum discrepancy between the quasienergies obtained from $U_{\bk}(T_d,0)$ and from the truncated Floquet matrix is $8.092\times10^{-5}\lambda$. The maximum absolute difference between the period-averaged spin polarization obtained by direct time integration and by the Fourier-component sum in Eq.~\eqref{eq:Sz_fourier_identity_SM} is $8.651\times10^{-5}$. Increasing $N_t$ or $N_F$ changes both observables by less than $10^{-4}$ according to Eqs.~\eqref{eq:quasienergy-error_SM} and \eqref{eq:spin-error_SM}; we therefore use $N_t=16$ and $N_F=10$ in the calculations reported above.

\section{Analytical time evolution of spin polarization}
\label{sec:analytics-time-evolution}

In this section, we use the Heisenberg equations of motion to clarify the appearance of momentum-even components in the spin polarization of $p$-wave magnets. We derive the time-dependent band spin polarization and determine the resonance condition under which the external driving frequency coincides with the interband energy difference.

We consider the spin operator per sector $\hat{\mathbf{S}} = \bm{\sigma}/2$. Its mean value is $\mathbf{S}(\mathbf{k},t) =\langle \hat{\mathbf{S}}\rangle$. In what follows, we call $\mathbf{S}(\mathbf{k},t)$ spin polarization.

The general form of the Heisenberg equation of motion for an operator $A(t)$ is
\begin{equation}
\label{ate-Heisenberg-gen}
i \partial_t A_H(t) = \left[A_H(t), H_H(t) \right] +i \left(\frac{\partial A_S}{\partial t}\right)_H.
\end{equation}
Here, the subscripts $H$ and $S$ denote the Heisenberg and Schr\"{o}dinger representations, respectively.

The equation of motion for the spin polarization reads
\begin{equation}
\label{ate-Heisenberg-S}
i  \partial_t \mathbf{S}(\bk,t) = \left[\mathbf{S}(\bk,t), H(t) \right] -i \frac{\mathbf{S} -\mathbf{S}_0}{\tau},
\end{equation}
where $\mathbf{S}_0$ is the equilibrium spin polarization, so only deviations from equilibrium are dissipated. We assume that the scattering time is isotropic and originates from scattering by pointlike impurities~\footnote{In general, the matrix element of the impurity potential requires more careful treatment because it can depend on momentum; see Ref.~\cite{Denisov-Zutic-AnisotropicSpinDynamics-2024} for a discussion in altermagnets.} This phenomenological approach does not distinguish between Dyakonov-Perel and Elliott-Yafet mechanisms.

To describe the electron quasiparticles, we use the continuum model of a $p$-wave magnet given in Eq.~\eqref{ate-H-def}.

In the presence of an electromagnetic field, we perform the following substitution:
\begin{equation}
\bk \to \bk +e \mathbf{A}(t),
\end{equation}
where $-e$ is the charge of the electron.

Substituting Eq.~\eqref{ate-H-def} into the Heisenberg equation \eqref{ate-Heisenberg-S}, we obtain the following evolution equations for the spin polarization:
\begin{eqnarray}
 \partial_t S_x &=& -2S_y M\left(\bk + e \mathbf{A}(t)\right) -\frac{S_x -S_{x,0}}{\tau},\\
 \partial_t S_y &=& 2 S_x M\left(\bk + e \mathbf{A}(t)\right) - 2 \eta S_z J_{sd} -\frac{S_y -S_{y,0}}{\tau}, \\
 \partial_t S_z &=& 2 \eta S_y J_{sd} -\frac{S_z -S_{z,0}}{\tau}.
\end{eqnarray}
In what follows, we use $M\left(\bk,t\right)=M\left(\bk + e\mathbf{A}(t)\right)$. In terms of the effective Zeeman field,
\begin{equation}
\label{ate-B-eff}
\mathbf{B}(\bk,t) = \left\{\eta J_{sd},0,M(\bk,t)\right\},
\end{equation}
we obtain the compact equation
\begin{equation}
\label{ate-S-eq}
\partial_t \mathbf{S}(\bk,t) = 2 \mathbf{B}(\bk,t)\times\mathbf{S}(\bk,t) -\frac{\mathbf{S}(\bk,t) -\mathbf{S}_0(\bk)}{\tau}.
\end{equation}

Assuming a weak external field, we expand
\begin{equation}
M(\bk,t) \equiv M\left(\bk + e \mathbf{A}(t)\right) \approx M(\bk) + e \mathbf{A}(t) \cdot \bm{\nabla}_{\bk} M(\bk).
\label{ate-app-M-exp_SM}
\end{equation}
The first and second terms in this expression are odd and even in momentum $\bk$, respectively. The solution to Eq.~\eqref{ate-S-eq} can then be sought as $\mathbf{S}(\bk,t) = \mathbf{S}_0(\bk) +\delta \mathbf{S}(\bk,t)$, with $\delta \mathbf{S}(\bk,t)$ determined by the external field, $\delta \mathbf{S}(\bk,t) \sim \mathbf{A}(t)$.

Then, Eq.~\eqref{ate-S-eq} is rewritten as
\begin{equation}
\label{ate-S-eq-fin}
D_t \delta \mathbf{S}(\bk,t) = 2 \mathbf{B}(\bk)\times \delta\mathbf{S}(\bk,t) +2 \delta \mathbf{B}(\bk,t)\times \mathbf{S}_0(\bk),
\end{equation}
where $D_t = \partial_t +1/\tau$,  $\mathbf{B}(\bk) = \left\{\eta J_{sd},0,M(\bk)\right\}$, $\delta \mathbf{B}(\bk, t) = \left\{0,0,\delta M(\bk,t)\right\}$, and $\mathbf{S}_0(\bk) = s \left\{\eta J_{sd}, 0, M(\bk)\right\}/[2B(\bk)]$, where $s=\pm$ corresponds to the two bands of our continuum model.

We assume that the external field oscillates as $\mathbf{A}(t) = \mathbf{A}_0 e^{-i\omega t}$; therefore, $\delta \mathbf{S}(\bk,t) = \delta \mathbf{S}(\bk) e^{-i\omega t}$. It is convenient to introduce the following notations:
\begin{equation}
\gamma_\omega=\frac{1}{\tau} -i\omega, \quad D_\omega(\mathbf{k})=4B^2(\mathbf{k})+\gamma_\omega^2.
\label{ate-gamma-D-def}
\end{equation}
The driving term is
\begin{equation}
\delta\mathbf{B}(\mathbf{k},\omega)\times \mathbf{S}_0(\mathbf{k}) = \left\{0,\delta M(\mathbf{k},\omega)S_{x,0}(\mathbf{k}),0\right\}.
\end{equation}
Then, the components of Eq.~\eqref{ate-S-eq-fin} are
\begin{eqnarray}
\label{ate-dSx-component}
\gamma_\omega\delta S_x(\bk) &=& -2M(\bk) \delta S_y(\bk),\\
\label{ate-dSy-component}
\gamma_\omega\delta S_y(\bk) &=& 2 \left[M(\bk) \delta S_x(\bk) -\eta J_{sd}\delta S_z(\bk) +S_{x,0}(\bk)\delta M(\bk) \right],\\
\label{ate-dSz-component}
\gamma_\omega\delta S_z(\bk) &=& 2\eta J_{sd}\delta S_y(\bk).
\end{eqnarray}
From Eqs.~\eqref{ate-dSx-component} and \eqref{ate-dSz-component}, we obtain
\begin{equation}
\delta S_x(\bk) = -\frac{M(\bk)}{\eta J_{sd}}\delta S_z(\bk), 
\quad 
\delta S_y(\bk) = \frac{\gamma_\omega}{2\eta J_{sd}}\delta S_z(\bk).
\label{ate-dS-relations}
\end{equation}
Substituting these relations into Eqs.~\eqref{ate-dSx-component}--\eqref{ate-dSz-component}, we obtain
\begin{eqnarray}
\label{ate-dSx-solution}
\delta S_x(\mathbf{k}) &=& -\frac{4M(\mathbf{k})S_{x,0}(\mathbf{k})} {D_\omega(\mathbf{k})} \delta M(\mathbf{k}) = -\frac{4\eta J_{sd}S_{z,0}(\mathbf{k})}{D_\omega(\mathbf{k})} \delta M(\mathbf{k}) = -s\eta \frac{2M(\mathbf{k})J_{sd}}{B(\mathbf{k})D_\omega(\mathbf{k})} \delta M(\mathbf{k}),\\
\label{ate-dSy-solution}
\delta S_y(\mathbf{k}) &=& \frac{2\gamma_\omega S_{x,0}(\mathbf{k})}
{D_\omega(\mathbf{k})} \delta M(\mathbf{k}) = \frac{2\eta J_{sd}\gamma_\omega S_{z,0}(\mathbf{k})}{M(\mathbf{k})D_\omega(\mathbf{k})} \delta M(\mathbf{k}) = s\eta \frac{J_{sd}\gamma_\omega} {B(\mathbf{k})D_\omega(\mathbf{k})} \delta M(\mathbf{k}),\\
\label{ate-dSz-solution}
\delta S_z(\mathbf{k}) &=& \frac{4\eta J_{sd}S_{x,0}(\mathbf{k})} {D_\omega(\mathbf{k})} \delta M(\mathbf{k}) = \frac{4J_{sd}^2S_{z,0}(\mathbf{k})} {M(\mathbf{k})D_\omega(\mathbf{k})} \delta M(\mathbf{k}) = s \frac{2 J_{sd}^2} {B(\mathbf{k})D_\omega(\mathbf{k})} \delta M(\mathbf{k}).
\end{eqnarray}

As one can see from Eqs.~\eqref{ate-dSx-solution} and \eqref{ate-dSy-solution}, there are $x$ and $y$ components of the spin polarization, which, however, vanish after summing over two sectors of the model. The $z$-component $\delta S_z(\mathbf{k})$ survives, hence it will be the main focus of the remainder of this section. 

By using Eq.~\eqref{ate-app-M-exp_SM}, $\mathbf{E}(t) = -\partial_t \mathbf{A}(t)$, and the relation for the Fourier transform  $\mathbf{E}_0 = i\omega \mathbf{A}_0$, we obtain
\begin{equation}
\delta M(\mathbf{k}) = \frac{1}{i\omega} e\mathbf{E}_0\cdot \nabla_{\mathbf{k}}M(\mathbf{k}).
\label{ate-deltaM-E}
\end{equation}
This allows us to rewrite Eq.~\eqref{ate-dSz-solution} as
\begin{eqnarray}
\label{ate-dis1-tau-app-Sz-harmonic-omega}
\sum_{\eta=\pm}\delta S_z(\bk) = 8 J_{sd}^2  S_{z,0}(\bk)  \frac{1}{i\omega}\frac{e\mathbf{E}_0 \cdot \bm{\nabla}_{\bk}M(\bk)}{M(\bk)} \frac{1}{4\left[M^2(\bk) +J_{sd}^2\right] +\left(i\omega  -1/\tau\right)^2}.
\end{eqnarray}

Let us discuss how the obtained expression transforms under $\bk \to -\bk$. By definition, $S_{z,0}(\bk) = -S_{z,0}(-\bk)$ and $M(\bk) =-M(-\bk)$ in $p$-wave magnets. The momentum derivative of $M(\bk)$ is an even function. Therefore, the expression in Eq.~\eqref{ate-dis1-tau-app-Sz-harmonic-omega} is even in $\bk$.

Equation~\eqref{ate-dis1-tau-app-Sz-harmonic-omega} also exhibits resonances. In the limit $\tau\to\infty$, the resonances occur at
\begin{eqnarray}
\label{ate-exact-Sz-omega}
\omega_{\pm} = \pm 2\sqrt{M^2(\bk) +J_{sd}^2}.
\end{eqnarray}
The positive-frequency pole is $\epsilon_{+}-\epsilon_{-}$; see Eq.~\eqref{ate-eps}. Hence, we expect the system to show a resonance at $\omega = \epsilon_{+}-\epsilon_{-}$.

Thus, the exchange coupling term $J_{sd}$ allows an even component of the spin polarization to appear during the time evolution. This behavior differs qualitatively from that of simplified models of odd-parity magnets containing the $M(\bk) \sigma_z$ term but no $J_{sd} \sigma_x$ term; see, e.g., Refs.~\cite{Maeda-Tanaka:2024, Fu-Cayao-LightinducedFloquetSpintriplet-2026, Ezawa-TunnelingMagnetoresistanceJunction-2026, Ezawa-QuantumGeometryXwave-2026, Froldi-Freire-HighlyEfficientSuperconducting-2026}.

\section{Kubo formulas}
\label{sec:analytics-kubo}

In this section, we formulate the Kubo linear-response approach to the magnetization induced by a driving field. This approach takes into account both the intraband redistribution of charge carriers and the interband coherence. Unlike Refs.~\cite{Garate-MacDonald-InfluenceTransportCurrent-2009, Chakraborty-Sinova-HighlyEfficientNonrelativistic-2024}, we do not assume the dc limit of vanishing driving frequency.

\subsection{General formulas}
\label{sec:analytics-kubo-gen}

We start with the following generic expression in Matsubara-frequency space:
\begin{eqnarray}
\label{kubo-chiO-def}
\chi_{ij}^{(O)}(i\Omega_m) &=& -\frac{T}{i\Omega_m} \sum_{i\omega_n} \int \frac{d^dk}{(2\pi)^d} \mbox{Tr}{\left\{ O_i G(\bk, i \omega_n +i\Omega_m) J_j G(\bk, i \omega_n)\right\}} \nonumber\\
&=& -\frac{T}{i\Omega_m} \sum_{i\omega_n} \sum_{a b} \int \frac{d^dk}{(2\pi)^d} O_{i;ab} G_b(\bk, i \omega_n +i\Omega_m) J_{j;ba} G_a(\bk, i \omega_n) \nonumber\\
&=& -\frac{1}{i\Omega_m} \sum_{a b} \int \frac{d^dk}{(2\pi)^d} \frac{n_F(\epsilon_b) -n_F(\epsilon_a)}{i\Omega_m +\epsilon_a -\epsilon_b} O_{i;ab} J_{j;ba},
\end{eqnarray}
where $d$ is the spatial dimension. In the second line, we switched to the band basis using the Green function $G_a(\bk, i \omega_n) = 1/\left[i\omega_n -\epsilon_{a}\right]$; in the last line, we performed the Matsubara summation. We used~\footnote{This expression is straightforwardly derived by separating the simple fractions and using the fact that $i\Omega_m$ is a bosonic Matsubara frequency.}
\begin{equation}
\label{kubo-Matsubara}
T \sum_{i\omega_n} \frac{1}{i\omega_n -\epsilon_a} \frac{1}{i\omega_n +i\Omega_m -\epsilon_b} = \frac{n_F(\epsilon_b) -n_F(\epsilon_a)}{i\Omega_m +\epsilon_a -\epsilon_b},
\end{equation}
where $n_{F}(\epsilon) = 1/\left[ 1 +e^{(\epsilon -\mu)/T}\right]$ is the Fermi-Dirac distribution. In the above expressions, $a,b$ denote all bands of the system. In the model \eqref{ate-H-def}, the band index includes the sector index $\eta$. 

The expression \eqref{kubo-chiO-def} is valid for the spin and orbital Edelstein effects, with $O$ denoting the corresponding spin or orbital operator.
In writing Eq.~\eqref{kubo-chiO-def}, we used the velocity gauge. To avoid issues with the intraband limit in this gauge, it is more convenient to employ the length gauge. This can be done via the following replacement at $a\neq b$:
\begin{equation}
\label{kubo-length-gauge}
\frac{\mathbf{v}_{ab}}{i\Omega_m} \to \frac{1}{i} \frac{\mathbf{v}_{ab}}{\epsilon_a -\epsilon_b}.
\end{equation}
Here, the matrix elements of the velocity operator $\mathbf{v}_{ab}$ appear in the current operator as $\mathbf{J}_{ab} = -e \mathbf{v}_{ab}$.

Analytical continuation is performed as $i\Omega_m \to \Omega +i/\tau$. Then, the interband part of the Edelstein response tensor is:
\begin{eqnarray}
\label{kubo-chiO-1}
\chi_{ij}^{(O); {\rm inter}}(\Omega) = i e \sum_{a\neq b} \int \frac{d^dk}{(2\pi)^d} \frac{n_F(\epsilon_b) -n_F(\epsilon_a)}{\left(\epsilon_b -\epsilon_a\right) \left(\Omega +i/\tau +\epsilon_a -\epsilon_b\right)} O_{i;ab} v_{j;ba}.
\end{eqnarray}

The intraband term is derived by taking the limit $\epsilon_b\to \epsilon_a$:
\begin{eqnarray}
\label{kubo-chiO-intra}
\chi_{ij}^{(O); {\rm intra}}(\Omega) = -i e  \sum_a \int \frac{d^dk}{(2\pi)^d} \frac{-n_F'(\epsilon_a)}{\Omega +i/\tau} O_{i;aa} v_{j;aa}.
\end{eqnarray}
In the limit of vanishing temperature, $\lim_{T\to0} n_F(\epsilon) = \tf{\mu -\epsilon}$ and $\lim_{T\to0} n_F'(\epsilon) = -\df{\mu -\epsilon}$.

\subsection{Spin Edelstein response}
\label{sec:analytics-kubo-spin}

Let us consider the spin Edelstein response for the continuum model of a $p$-wave magnet given in Eq.~\eqref{ate-H-def}. We focus on the contribution of a single $\eta=+$ sector. The contribution from the other sector is obtained by $J_{sd} \to -J_{sd}$.

We start with the \emph{intraband} part:
\begin{eqnarray}
\label{kubo-chiO-intra-1}
\chi_{zx}^{(S); {\rm intra}}(\Omega) 
&=& -\frac{i}{\Omega +i/\tau} e  \sum_{s=\pm} \int_0^{2\pi} \frac{d\varphi}{2\pi} \int \frac{dk^2}{4\pi} \frac{\df{k_{s}^2 -k^2}}{|\partial \epsilon_{s}/\partial k^2|} \left[\frac{\alpha^3 k^2 \cos^2{\varphi}}{2\Delta_k^2} +s\frac{\alpha k^2 \cos^2{\varphi}}{2m\Delta_k} \right]
\nonumber\\
&=& -\frac{i}{\Omega +i/\tau} e  \sum_{s=\pm} \int_0^{2\pi} \frac{d\varphi}{2\pi} \frac{\tf{k^2_{s}}}{4\pi} 
\frac{2}{ \left[2t_0\Delta_k +s \alpha^2 \cos^2{\varphi}\right]} \left[\frac{\alpha^3 k_{s}^2 \cos^2{\varphi}}{2\Delta_{k}} +s t_0 \alpha k_{s}^2 \cos^2{\varphi}\right],
\end{eqnarray} 
where we used the shorthand notation $\Delta_k =\sqrt{J_{sd}^2 +\alpha^2k^2 \cos^2{\varphi}}$, defined $t_0=1/(2m)$, and assumed $\bm{\alpha} = \alpha \hat{\mathbf{x}}$.

Assuming $\alpha\to0$ and using $k_{s}^2 \approx (\mu -sJ_{sd})/t_0$, we arrive at
\begin{eqnarray}
\label{kubo-chiO-intra-1-small-alpha}
\chi_{zx}^{(S); {\rm intra}}(\Omega) &\approx& -\frac{i}{\Omega +i/\tau} e  \sum_{s=\pm} \int_0^{2\pi} \frac{d\varphi}{2\pi} \frac{\tf{\mu -s J_{sd}}}{4\pi} \frac{-s \alpha (\mu -s J_{sd}) \cos^2{\varphi}}{J_{sd} t_0} \nonumber\\
&=& -\tau \frac{1 +i\Omega \tau}{1 +(\Omega \tau)^2} e  
\frac{\alpha}{8\pi J_{sd} t_0} \sum_{s=\pm} s \tf{\mu -s J_{sd}} (\mu -s J_{sd}).
\end{eqnarray} 
As one can see, the intraband part of the spin Edelstein response tensor scales as $\alpha$ for small $\alpha$. Its real part has a regular Drude peak $\sim \tau /(\Omega^2\tau^2+1)$.

The \emph{interband} contribution reads:
\begin{eqnarray}
\label{kubo-chiO-inter-1}
\chi_{zx}^{(S); {\rm inter}}(\Omega) &=& i e \sum_{s_1\neq s_2} \int_0^{2\pi} \frac{d\varphi}{2\pi} \int \frac{dk^2}{4\pi} \frac{\tf{\mu -\epsilon_{s_2}} -\tf{\mu -\epsilon_{s_1}}}{(s_2-s_1)\Delta_k \left(\Omega +i/\tau -(s_2-s_1)\Delta_k\right)} \frac{\alpha J_{sd}^2}{2\Delta_k^2} \nonumber\\
&=& i e \sum_{s=\pm} \int_0^{2\pi} \frac{d\varphi}{2\pi} \int_{k_{+}^2 \tf{k_{+}^2}}^{k_{-}^2} \frac{dk^2}{4\pi} \frac{-1}{2\Delta_k \left(\Omega +i/\tau -2s\Delta_k\right)} \frac{\alpha J_{sd}^2}{2\Delta_k^2}.
\end{eqnarray}
Here, we took into account that $k_{-}^2>0$ for $\mu>0$, whereas $k_{+}^2$ may become negative, signaling that the upper band is empty. The integral over $k$ can be evaluated analytically. The final result is, however, cumbersome:
\begin{eqnarray}
\label{kubo-chiO-inter-int-Re-1}
&&\RE{i\int dk^2 \frac{\Omega +2s\Delta_k -i/\tau}{2\Delta_k \left((\Omega +2s\Delta_k)^2 +1/\tau^2\right)} \frac{\alpha J_{sd}^2}{2\Delta_k^2}} \nonumber\\
&&= -\frac{J_{sd}^2 \tau}{2\alpha \Delta_k (1 + \tau^2 \Omega^2)^2} \Big\{ 2 + 2 \tau^2 \Omega^2 + \sqrt{2} \tau \sqrt{2 J_{sd}^2 + k^2 \alpha^2 + k^2 \alpha^2 \cos(2\varphi)} \big[ \arctan(\tau (2\Delta_k - \Omega)) + \arctan(\tau (2\Delta_k + \Omega)) \nonumber\\
&&+ \tau \Omega \big(\tau \Omega (\arctan(\tau (-2\Delta_k + \Omega)) - \arctan(\tau (2\Delta_k + \Omega))) + \ln(1 + \tau^2 (4 J_{sd}^2 - 4 \Delta_k \Omega + \Omega^2) + 4 k^2 \alpha^2 \tau^2 \cos^2(\varphi)) \nonumber\\
&&- \ln(1 + \tau^2 (4 J_{sd}^2 + 4 \Delta_k \Omega + \Omega^2) + 4 k^2 \alpha^2 \tau^2 \cos^2(\varphi)) \big) \big] \Big\} \sec^2(\varphi) \Bigg|_{k_{+}^2 \tf{k_{+}^2}}^{k_{-}^2}
\end{eqnarray}
and
\begin{eqnarray}
\label{kubo-chiO-inter-int-Im-1}
&&\IM{i\int dk^2 \frac{\Omega +2s\Delta_k -i/\tau}{2\Delta_k \left((\Omega +2s\Delta_k)^2 +1/\tau^2\right)} \frac{\alpha J_{sd}^2}{2\Delta_k^2}} \nonumber\\
&&= -\frac{J_{sd}^2 \tau^2}{4 \alpha \Delta_k (1 + \tau^2 \Omega^2)^2} \Big\{ 4(\Omega + \tau^2 \Omega^3) - 8\Delta_k \tau \Omega \arctan(\tau(-2\Delta_k + \Omega)) + 8\Delta_k \tau \Omega \arctan(\tau(2\Delta_k + \Omega)) \nonumber\\
&&+ \sqrt{2} (-1 + \tau^2 \Omega^2) \sqrt{2J_{sd}^2 + k^2 \alpha^2 + k^2 \alpha^2 \cos(2\varphi)} \big[ \ln(1 + \tau^2(4J_{sd}^2 - 4\Delta_k \Omega + \Omega^2) + 4k^2 \alpha^2 \tau^2 \cos^2(\varphi)) \nonumber\\
&&- \ln(1 + \tau^2(4J_{sd}^2 + 4\Delta_k \Omega + \Omega^2) + 4k^2 \alpha^2 \tau^2 \cos^2(\varphi)) \big] \Big\} \sec^2(\varphi) \Bigg|_{k_{+}^2 \tf{k_{+}^2}}^{k_{-}^2}.
\end{eqnarray}

Using Eqs.~\eqref{kubo-chiO-inter-int-Re-1} and \eqref{kubo-chiO-inter-int-Im-1}, assuming $\mu>J_{sd}$, and expanding to leading order in $\alpha$, we simplify Eq.~\eqref{kubo-chiO-inter-1} as
\begin{eqnarray}
\label{kubo-chiO-inter-app-1}
\chi_{zx}^{(S); {\rm inter}}(\Omega) &\approx& -\frac{e}{4\pi} \int_0^{2\pi} \frac{d\varphi}{2\pi}  
2 \alpha \tau \frac{(1 + 4 J_{sd}^2 \tau^2 + \tau^2 \Omega^2) +i\tau \Omega (1 - 4 J_{sd}^2 \tau^2 + \tau^2 \Omega^2)}{t_0 \left[ 16 J_{sd}^4 \tau^4 + (1 + \tau^2 \Omega^2)^2 + 8J_{sd}^2\tau^2(1 - \tau^2 \Omega^2) \right]} \nonumber\\
&=& i\frac{e}{2\pi} \frac{\alpha}{t_0} \frac{\Omega +i/\tau}{(\Omega +i/\tau)^2 -4J_{sd}^2} 
\stackrel{\tau \to \infty}{\approx} -\frac{e}{2\pi} \frac{\alpha}{t_0} \left[\frac{1}{\tau} \frac{\Omega^2 +4J_{sd}^2}{\left(\Omega^2 -4J_{sd}^2\right)^2} 
+\frac{i\Omega }{\Omega^2 -4J_{sd}^2}\right].
\end{eqnarray}
Similar to the intraband part, the interband part scales as $\alpha$. The resonant behavior at $\Omega =2J_{sd}$ is also evident from Eq.~\eqref{kubo-chiO-inter-app-1}. The intraband and interband expressions in Eqs.~\eqref{kubo-chiO-intra-1-small-alpha} and \eqref{kubo-chiO-inter-app-1} correspond to Eqs.~(8) and (9), respectively, in the main text.

\begin{figure*}[!ht]
\centering
\includegraphics[width=0.45\textwidth]{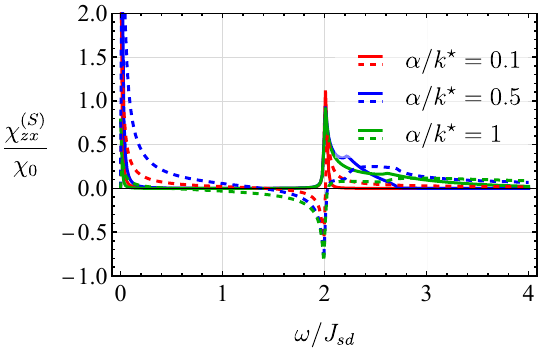}
\caption{
The spin Edelstein response coefficient $\chi_{zx}$ as a function of the driving frequency $\omega$. Solid and dashed lines correspond to the real and imaginary parts of $\chi_{zx}$, respectively. We used the Kubo formulas \eqref{kubo-chiO-intra-1-small-alpha} and \eqref{kubo-chiO-inter-app-1}, where we summed over the sectors. In all panels, we use $k^{\star}=\sqrt{2m J_{sd}}$ and fix $\mu=2\,J_{sd}$ and $\tau = 100\,J_{sd}^{-1}$. The normalization coefficient is $\chi_0=e \sqrt{2m/J_{sd}}$.
}
\label{fig:kubo-chizx}
\end{figure*}

The results for the spin Edelstein effect for the effective model \eqref{ate-H-def} are shown in Fig.~\ref{fig:kubo-chizx}.

\subsection{Orbital Edelstein effect}
\label{sec:analytics-kubo-orbital}

For the orbital Edelstein effect, the orbital angular momentum $\mathbf{L}$ can be calculated as $\mathbf{L} = -(1/g_L \mu_B) \mathbf{m}$, where $g_L=1$ is the Land\'{e} $g$-factor, $\mu_B = e/(2m_e)$ is the Bohr magneton, and $\mathbf{m}$ is the orbital magnetic moment.

The dynamical orbital Edelstein effect is calculated via Eqs.~\eqref{kubo-chiO-1} and \eqref{kubo-chiO-intra} with $\mathbf{O} = \mathbf{L}$. We have:
\begin{eqnarray}
\label{kubo-orb-chiO-inter}
\chi_{ij}^{(L); {\rm inter}}(\Omega) &=& i e \sum_{a\neq b} \int \frac{d^dk}{(2\pi)^d} \frac{n_F(\epsilon_b) -n_F(\epsilon_a)}{\left(\epsilon_b -\epsilon_a\right) \left(\Omega +i/\tau -\epsilon_b +\epsilon_a\right)} L_{i;ab} v_{j;ba},\\
\label{kubo-orb-chiO-intra}
\chi_{ij}^{(L); {\rm intra}}(\Omega) &=& i e  \sum_a \int \frac{d^dk}{(2\pi)^d} \frac{-n_F'(\epsilon_a)}{\Omega +i/\tau} L_{i;aa} v_{j;aa}.
\end{eqnarray}

The matrix elements of the velocity operator are:
\begin{equation}
\label{kubo-orb-vnn1-def}
\mathbf{v}_{nn'} =  \langle n | (\partial_{\bk}H) | n' \rangle = 
\partial_{\bk}\epsilon_n \delta_{nn'} +i \mathcal{\bm{A}}_{nn'} (\epsilon_n -\epsilon_{n'}),
\end{equation}
where $\mathcal{\bm{A}}_{nn'} = i \langle n | \partial_{\bk} | n' \rangle$ is the interband Berry connection. 

In the model given in Eq.~\eqref{ate-H-def}, there are four bands: two bands per sector. We assume that $n=1,2$ correspond to the lower and upper bands with $\eta=+1$, and $n=3,4$ correspond to the lower and upper bands with $\eta=-1$. The Berry connection reads:
\begin{eqnarray}
\label{kubo-Ann}
\mathcal{\bm{A}}_{11} &=& \mathcal{\bm{A}}_{22} = \mathbf{0},\\
\label{kubo-Ann1}
\mathcal{\bm{A}}_{12} &=& -\mathcal{\bm{A}}_{21} = \frac{i J_{sd} \bm{\alpha}}{2\left[(\bm{\alpha}\cdot \mathbf{k})^2 +J_{sd}^2\right]}.
\end{eqnarray}
Therefore, the matrix elements of the velocity operator given in Eq.~\eqref{kubo-orb-chiO-inter} in the band basis are:
\begin{eqnarray}
\label{kubo-vnn}
\mathbf{v}_{11} &=& \frac{\bk}{m} -\frac{\bm{\alpha} (\bm{\alpha}\cdot \mathbf{k})}{\sqrt{(\bm{\alpha}\cdot \mathbf{k})^2 +J_{sd}^2}},\\
\label{kubo-v22}
\mathbf{v}_{22} &=& \frac{\bk}{m} +\frac{\bm{\alpha} (\bm{\alpha}\cdot \mathbf{k})}{\sqrt{(\bm{\alpha}\cdot \mathbf{k})^2 +J_{sd}^2}},\\
\label{kubo-vnn1}
\mathbf{v}_{12} &=& \mathbf{v}_{21} = \frac{J_{sd} \bm{\alpha}}{\sqrt{(\bm{\alpha}\cdot \mathbf{k})^2 +J_{sd}^2}}.
\end{eqnarray}
The results for the other sector with $\eta =-1$ are obtained by replacing $J_{sd} \to -J_{sd}$.

The matrix elements of the orbital magnetic moment contain a diagonal (Abelian) part and off-diagonal (non-Abelian) elements. According to Ref.~\cite{Cysne-Rappoport-DescriptionOrbitalHall-2025}, one has:
\begin{eqnarray}
\label{kubo-mnn1-def}
\mathbf{m}_{nn'} &=& \mathbf{m}_{nn'}^{(0)} + \mathbf{m}_{nn'}^{(1)} +\mathbf{m}_{nn'}^{(2)},\\
\label{kubo-mnn1-0-def}
\mathbf{m}_{nn'}^{(0)} &=& -\frac{ie}{2} \left\langle \partial_{\bk} n \right| \times \left[H(\bk) - \frac{1}{2}(\epsilon_{n} +\epsilon_{n'})\right] \left| \partial_{\bk} n'\right\rangle,\\
\label{kubo-mnn1-1-def}
\mathbf{m}_{nn'}^{(1)} &=& \frac{e}{4} \left[\left(\mathcal{\bm{A}}_{nn} +\mathcal{\bm{A}}_{n'n'}\right) \times \mathbf{v}_{nn'} \right] (1-\delta_{nn'}),\\
\label{kubo-mnn1-2-def}
\mathbf{m}_{nn'}^{(2)} &=& \frac{e}{4} \left[\left(\mathbf{v}_{nn} +\mathbf{v}_{n'n'}\right) \times \mathcal{\bm{A}}_{nn'} \right] (1-\delta_{nn'}).
\end{eqnarray}
Here, $\mathbf{m}_{nn'}^{(1)}$ and $\mathbf{m}_{nn'}^{(2)}$ are the off-diagonal corrections. The matrix elements of the orbital angular momentum follow from the relation $\mathbf{L} = -(1/g_L \mu_B) \mathbf{m}$.

Using the expressions in Eqs.~\eqref{kubo-mnn1-def}--\eqref{kubo-mnn1-2-def}, we obtain the following orbital magnetic-moment matrix elements:
\begin{eqnarray}
\label{kubo-mnn-fin}
\mathbf{m}_{nn} &=&\mathbf{0},\\
\label{kubo-mnn1-fin}
\mathbf{m}_{12} &=& -\mathbf{m}_{21} = -\frac{ie J_{sd}}{4m} \frac{\bm{\alpha} \times \bk}{J_{sd}^2 +(\bm{\alpha}\cdot \mathbf{k})^2}.
\end{eqnarray}

Therefore, while $L_{nn}v_{nn}=0$, the combination $L_{12}v_{21}$, given by Eqs.~\eqref{kubo-vnn1} and \eqref{kubo-mnn1-fin}, is odd in $\bk$. According to Eqs.~\eqref{kubo-orb-chiO-inter} and \eqref{kubo-orb-chiO-intra}, the momentum-integrated linear orbital Edelstein response therefore vanishes. A nonzero response can, however, appear at higher order.

\section{Density matrix formalism}
\label{sec:density-matrix}

In this section, we present the details of the density matrix formalism~\cite{Sipe-Shkrebtii:2000, Vasko-Raichev:book, Fregoso-BulkPhotospinEffect-2022} used in the main text to show how the even-parity orbital magnetization emerges in a driven $p$-wave magnet. 

For a fixed sector, the continuum model in Eq.~\eqref{ate-H-def} is a two-band model. Therefore, we first formulate a general theory for an arbitrary two-band Hamiltonian in Sec.~\ref{sec:density-matrix-gen}. We then apply this theory to the continuum model in Eq.~\eqref{ate-H-def} in Sec.~\ref{sec:density-matrix-pwave}.

\subsection{General formalism}
\label{sec:density-matrix-gen}

\subsubsection{Preliminaries and zero-order results}
\label{sec:density-matrix-0}

We consider the following two-band Hamiltonian:
\begin{equation}
H_0(\mathbf{k}) = d_0(\mathbf{k})\sigma_0 + \mathbf{d}(\mathbf{k})\cdot\boldsymbol{\sigma}.
\label{dm-0-H-def_SM}
\end{equation}
A spatially uniform electric field is introduced in the length gauge as
\begin{equation}
H_E(t) = e\,\mathbf{E}(t)\cdot \mathbf{r},
\end{equation}
where $-e$ with $e>0$ is the electron charge and $\mathbf{r}$ is the coordinate operator. Equivalently, in the semiclassical representation, the electric field produces the drift term
\begin{equation}
\dot{\mathbf{k}} = -e\mathbf{E}(t).
\end{equation}

The equation of motion (quantum Liouville equation) for the density matrix is
\begin{equation}
\frac{\partial \rho_{\mathbf{k}}}{\partial t} - e\mathbf{E}(t)\cdot\nabla_{\mathbf{k}}\rho_{\mathbf{k}} +i \left[ H_0(\mathbf{k}), \rho_{\mathbf{k}}\right] = I_{\rm coll}[\rho_{\mathbf{k}}].
\label{dm-0-eq-def_SM}
\end{equation}
In the relaxation-time approximation, the collision integral is
\begin{equation}
I_{\rm coll}[\rho_{\mathbf{k}}] = - \frac{\rho_{\mathbf{k}} -\rho^{(0)}_{\mathbf{k}}}{\tau}.
\label{dm-0-Icoll-def_SM}
\end{equation}
Here, $\tau$ is the relaxation time, and $\rho^{(0)}_{\mathbf{k}}$ is the density matrix in the absence of electric fields.
Therefore,
\begin{equation}
\frac{\partial \rho_{\mathbf{k}}}{\partial t} - e\mathbf{E}(t)\cdot\nabla_{\mathbf{k}}\rho_{\mathbf{k}} + i\left[\mathbf{d}(\mathbf{k})\cdot\boldsymbol{\sigma}, \rho_{\mathbf{k}}\right] = - \frac{\rho_{\mathbf{k}} - \rho^{(0)}_{\mathbf{k}}}{\tau}.
\label{dm-0-rho-eom-generic_SM}
\end{equation}

It is useful to decompose the density matrix as
\begin{equation}
\label{eq:dm-0-rho-decomp}
\rho_{\mathbf{k}} = \frac{1}{2}\left[n_{\mathbf{k}}\sigma_0 + \mathbf{s}_{\mathbf{k}}\cdot\boldsymbol{\sigma} \right].
\end{equation}
Using
\begin{equation}
i\left[\mathbf{d}\cdot\boldsymbol{\sigma}, \rho_{\mathbf{k}}\right] = -\left[\mathbf{d}\times\mathbf{s}_{\mathbf{k}}\right]\cdot\boldsymbol{\sigma},
\end{equation}
we obtain
\begin{eqnarray}
\label{eq:dm-0-n-eom-generic_SM}
&&\frac{\partial n_{\mathbf{k}}}{\partial t} - e\mathbf{E}(t)\cdot\nabla_{\mathbf{k}}n_{\mathbf{k}} = -\frac{n_{\mathbf{k}} -n^{(0)}_{\mathbf{k}}}{\tau},\\
\label{eq:dm-0-s-eom-generic_SM}
&&\frac{\partial \mathbf{s}_{\mathbf{k}}}{\partial t} -e\mathbf{E}(t)\cdot\nabla_{\mathbf{k}}\mathbf{s}_{\mathbf{k}} = 2\mathbf{d}(\mathbf{k})\times\mathbf{s}_{\mathbf{k}} -\frac{\mathbf{s}_{\mathbf{k}} -\mathbf{s}^{(0)}_{\mathbf{k}}}{\tau}.
\end{eqnarray}

The equilibrium density matrix can be decomposed as
\begin{equation}
\rho^{(0)}_{\mathbf{k}} = f_+(\mathbf{k})P_+(\mathbf{k}) + f_-(\mathbf{k})P_-(\mathbf{k}),
\label{eq:dm-0-rho-0-decomp}
\end{equation}
where
\begin{equation}
P_\pm(\mathbf{k}) = \frac{1}{2} \left[ \sigma_0 \pm \hat{\mathbf{d}}(\mathbf{k})\cdot\boldsymbol{\sigma} \right], \qquad \epsilon_\pm(\mathbf{k}) = d_0(\mathbf{k}) \pm d(\mathbf{k}).
\end{equation}
Here, $d(\mathbf{k})=|\mathbf{d}(\mathbf{k})|$ and $\hat{\mathbf{d}}(\mathbf{k}) = \mathbf{d}(\mathbf{k})/d(\mathbf{k})$. Thus,
\begin{equation}
n^{(0)}_{\mathbf{k}} = f_+(\mathbf{k}) + f_-(\mathbf{k}), \qquad \mathbf{s}^{(0)}_{\mathbf{k}} = 
\Delta f(\mathbf{k}) \hat{\mathbf{d}}(\mathbf{k}),
\label{eq:dm-0-rho-n-s}
\end{equation}
where $\Delta f(\mathbf{k}) = f_+(\mathbf{k}) -f_-(\mathbf{k})$ and $f_{s}(\bk) = n_{F}(\epsilon_{s,\bk})$ is the equilibrium distribution function.

In what follows, we solve Eqs.~\eqref{dm-0-rho-eom-generic_SM}, \eqref{eq:dm-0-n-eom-generic_SM}, and \eqref{eq:dm-0-s-eom-generic_SM} perturbatively by writing $\rho_{\mathbf{k}} = \sum_{n=0}\rho^{(n)}_{\mathbf{k}}$. We decompose each contribution as
\begin{equation}
\label{eq:dm-0-rho-decomp-n}
\rho^{(n)}_{\mathbf{k}} = \frac{1}{2} \left[ n^{(n)}_{\mathbf{k}}\sigma_0 + \mathbf{s}^{(n)}_{\mathbf{k}} \cdot \boldsymbol{\sigma} \right].
\end{equation}

\subsubsection{First-order results}
\label{sec:density-matrix-1}

We now solve the density-matrix equation \eqref{dm-0-rho-eom-generic_SM} to first order in the electric field. The linearized kinetic equation is
\begin{equation}
\frac{\partial \rho_{\mathbf{k}}^{(1)}}{\partial t} +i\left[\mathbf{d}(\mathbf{k})\cdot\boldsymbol{\sigma}, \rho_{\mathbf{k}}^{(1)}\right] + \frac{\rho_{\mathbf{k}}^{(1)}}{\tau} = e\mathbf{E}(t)\cdot \nabla_{\mathbf{k}}\rho^{(0)}_{\mathbf{k}}.
\label{eq:dm-1-linear-rho-eom-general}
\end{equation}
We use the standard Fourier representations $\mathbf{E}(t) = \mathbf{E}_{\omega}e^{-i\omega t}$~\footnote{The physical field also contains the complex conjugate component.} and $ \rho_{\mathbf{k}}^{(1)}(t) = \rho_{\mathbf{k},\omega}^{(1)}e^{-i\omega t}$ and define $\gamma_{\omega} = 1/\tau - i\omega$. Then, Eq.~\eqref{eq:dm-1-linear-rho-eom-general} becomes
\begin{equation}
\gamma_{\omega} \rho_{\mathbf{k},\omega}^{(1)} + i\left[\mathbf{d}(\mathbf{k})\cdot\boldsymbol{\sigma}, \rho_{\mathbf{k},\omega}^{(1)}\right] = e\mathbf{E}_{\omega} \cdot \nabla_{\mathbf{k}} \rho^{(0)}_{\mathbf{k}}.
\label{eq:dm-1-linear-rho-eom-frequency}
\end{equation}

We decompose the components of the equilibrium and nonequilibrium density matrices according to Eq.~\eqref{eq:dm-0-rho-decomp-n} as
\begin{equation}
\rho^{(0)}_{\mathbf{k}} = \frac{1}{2}\left[n^{(0)}_{\mathbf{k}}\sigma_0 +\mathbf{s}^{(0)}_{\mathbf{k}}\cdot\boldsymbol{\sigma}\right], 
\qquad \rho_{\mathbf{k},\omega}^{(1)} = \frac{1}{2} \left[n_{\mathbf{k},\omega}^{(1)}\sigma_0 + \mathbf{s}_{\mathbf{k},\omega}^{(1)}\cdot\boldsymbol{\sigma}\right].
\end{equation}
The equilibrium density matrix is given in Eqs.~\eqref{eq:dm-0-rho-0-decomp} and \eqref{eq:dm-0-rho-n-s}. Substituting this decomposition into Eq.~\eqref{eq:dm-1-linear-rho-eom-frequency}, we obtain
\begin{eqnarray}
\label{eq:dm-1-delta-n-general}
&&\gamma_{\omega} n_{\mathbf{k},\omega}^{(1)} = eE^a_{\omega} \partial_{k_a} n^{(0)}_{\mathbf{k}}, \nonumber\\
\label{eq:dm-1-delta-s-linear-eom-general}
&&\gamma_{\omega} \mathbf{s}_{\mathbf{k},\omega}^{(1)} - 2\left[\mathbf{d}(\mathbf{k})\times\mathbf{s}_{\mathbf{k},\omega}^{(1)}\right] = eE^a_{\omega} \partial_{k_a} \mathbf{s}^{(0)}_{\mathbf{k}}.
\end{eqnarray}
The scalar correction is therefore
\begin{equation}
n_{\mathbf{k},\omega}^{(1)} = \frac{eE^a_{\omega} \partial_{k_a} \left[ f_+(\mathbf{k}) + f_-(\mathbf{k})\right]}{\gamma_{\omega}}.
\label{eq:dm-1-delta-n-solution-general}
\end{equation}

The right-hand side of Eq.~\eqref{eq:dm-1-delta-s-linear-eom-general} is
\begin{equation}
eE^a_{\omega} \partial_{k_a} \mathbf{s}^{(0)}_{\mathbf{k}} = eE^a_{\omega} \left[\left(\partial_{k_a}\Delta f\right)\hat{\mathbf{d}}  +\Delta f\partial_{k_a}\hat{\mathbf{d}}\right].
\end{equation}
The first term is parallel to $\hat{\mathbf{d}}$, whereas the second term is transverse because $\hat{\mathbf{d}}\cdot \partial_{k_a}\hat{\mathbf{d}} =0$.

To solve Eq.~\eqref{eq:dm-1-delta-s-linear-eom-general}, it is useful to introduce the inverse pseudospin propagator. For any vector source $\mathbf{B}$, define
\begin{equation}
\mathcal{R}_{\Omega}[\mathbf{B}] = \frac{\mathbf{B}_{\parallel}}{\gamma_{\Omega}} + \frac{\gamma_{\Omega}\mathbf{B}_{\perp}+ 2\left[\mathbf{d}\times\mathbf{B}_{\perp}\right]}{\gamma_{\Omega}^2+4d^2},
\label{eq:dm-1-R-def-general}
\end{equation}
where $\mathbf{B}_{\parallel} = \left( \mathbf{B}\cdot\hat{\mathbf{d}}\right) \hat{\mathbf{d}}$ and $\mathbf{B}_{\perp} = \mathbf{B} - \mathbf{B}_{\parallel}$.

The operator $\mathcal{R}_{\Omega}[\mathbf{B}]$ solves the following equation:
\begin{equation}
\gamma_{\Omega}\mathbf{x} - 2\left[\mathbf{d}\times\mathbf{x}\right] = \mathbf{B}, \qquad \mathbf{x} = \mathcal{R}_{\Omega}[\mathbf{B}],
\end{equation}
cf. Eq.~\eqref{eq:dm-1-delta-s-linear-eom-general}. Therefore, the first-order pseudospin correction can be written as
\begin{equation}
\mathbf{s}^{(1)}_{\mathbf{k},\omega} = eE^b_{\omega} \boldsymbol{\Phi}_{b}(\mathbf{k},\omega),
\label{eq:dm-1-delta-s-solution-general}
\end{equation}
where
\begin{equation}
\boldsymbol{\Phi}_{b}(\mathbf{k},\omega) = \frac{ \left( \partial_{k_b}\Delta f\right)\hat{\mathbf{d}}}{\gamma_{\omega}} + \Delta f \frac{\gamma_{\omega} \partial_{k_b}\hat{\mathbf{d}}  + 2\left[\mathbf{d}\times\partial_{k_b}\hat{\mathbf{d}}\right]}{\gamma_{\omega}^2+4d^2}.
\label{eq:dm-1-Phi-general}
\end{equation}

\subsubsection{Second-order response}
\label{sec:density-matrix-2}

The second-order density matrix obeys
\begin{equation}
\gamma_{\Omega} \rho^{(2)}_{\mathbf{k},\Omega} + i\left[\mathbf{d}(\mathbf{k})\cdot\boldsymbol{\sigma}, \rho^{(2)}_{\mathbf{k},\Omega} \right] = e \sum_{\omega_1+\omega_2=\Omega} E^a_{\omega_1} \partial_{k_a} \rho^{(1)}_{\mathbf{k},\omega_2}.
\label{eq:dm-2-rho-second-order-eom}
\end{equation}
The right-hand side accommodates both the rectified response at $\Omega=0$ and ac responses at $\Omega\neq0$.

Substituting Eq.~\eqref{eq:dm-1-delta-n-solution-general} into the scalar part of Eq.~\eqref{eq:dm-2-rho-second-order-eom}, we find
\begin{equation}
n^{(2)}_{\mathbf{k},\Omega} = e^2 \sum_{\omega_1+\omega_2=\Omega} E^a_{\omega_1} E^b_{\omega_2} \frac{\partial_{k_a}\partial_{k_b}
\left(f_+ + f_-\right)}{\gamma_{\Omega}\gamma_{\omega_2}}.
\label{eq:dm-2-n-second-general}
\end{equation}
Similarly, the pseudospin part of Eq.~\eqref{eq:dm-2-rho-second-order-eom} is
\begin{equation}
\mathbf{s}^{(2)}_{\mathbf{k},\Omega} = e^2 \sum_{\omega_1+\omega_2=\Omega} E^a_{\omega_1} E^b_{\omega_2} \mathcal{R}_{\Omega} \left[ \partial_{k_a} \boldsymbol{\Phi}_{b}(\mathbf{k},\omega_2) \right].
\label{eq:dm-2-s-second-general}
\end{equation}

\subsubsection{Orbital magnetization}
\label{sec:density-matrix-orbital}

The orbital magnetization is defined as:
\begin{equation}
\label{eq:dm-m-def}
\mathbf{M}(t) = \sum_{\eta} \int \frac{d^dk}{(2\pi)^d} \, \Tr\left\{\hat {\mathbf{m}}^{\rm band}_{\eta}(\mathbf{k})\rho^{\rm band}_{\eta,\mathbf{k}}(t) \right\}.
\end{equation}
Because the orbital-moment matrix elements are given in the band basis, see, e.g., Sec.~\ref{sec:analytics-kubo-orbital}, it is convenient to express the density matrix in the same basis. 

The transformation from the fixed $\sigma$ basis to the band $\lambda$ basis is performed via the unitary transformation
\begin{equation}
\rho^{\rm band}_{\eta,\mathbf{k}} = U_{\eta}^{\dag} \rho_{\eta,\mathbf{k}} U_{\eta},
\label{eq:dm-m-unitary}
\end{equation}
where $U_{\eta}$ is the matrix that diagonalizes the unperturbed Hamiltonian.

We provide explicit expressions for the continuum model of a $p$-wave magnet in Sec.~\ref{sec:density-matrix-pwave}.

\subsection{Continuum model of a $p$-wave magnet}
\label{sec:density-matrix-pwave}

Let us apply the general formalism developed in Sec.~\ref{sec:density-matrix-gen} to the continuum model of a $p$-wave magnet given in Eq.~\eqref{ate-H-def}.

\subsubsection{Zero order}
\label{sec:dm-0-pwave}

For the model given in Eq.~\eqref{ate-H-def}, we have
\begin{equation}
\label{eq:dm-0-pwave-d}
d_0(\bk) = \xi_{\mathbf{k}}, \qquad  \mathbf{d}_\eta(\mathbf{k})  = \left\{\eta J_{sd}, 0, M(\mathbf{k})\right\}, \qquad d_\eta(\mathbf{k}) = \sqrt{J_{sd}^2 + M^2(\mathbf{k})}.
\end{equation}

The equilibrium solutions for $n^{(0)}_{\eta,\mathbf{k}}$ and $\mathbf{s}^{(0)}_{\eta,\mathbf{k}}$ are
\begin{eqnarray}
\label{eq:dm-0-n0-model}
n^{(0)}_{\eta,\mathbf{k}} &=& f_{\eta,+}(\mathbf{k}) +f_{\eta,-}(\mathbf{k}),\\
\label{eq:dm-0-s0-model}
\mathbf{s}^{(0)}_{\eta,\mathbf{k}} &=& \frac{\Delta f_{\eta}(\mathbf{k})}{\sqrt{J_{sd}^2 + M^2(\mathbf{k})}} \left\{\eta J_{sd}, 0, M(\mathbf{k}) \right\},
\end{eqnarray}
where $\Delta f_{\eta}(\mathbf{k}) = f_{\eta,+}(\mathbf{k}) -f_{\eta,-}(\mathbf{k})$ and we added the subscript $\eta$ to keep track of the sector index.

\subsubsection{First order}
\label{sec:dm-1-pwave}

We now consider the first-order expressions. For the continuum model with $\mathbf{d}$ given in Eq.~\eqref{eq:dm-0-pwave-d}, we have
\begin{equation}
\partial_{k_a}\hat{\mathbf{d}}_{\eta} = \frac{\partial_{k_a}M}{d^3} \left\{-\eta J_{sd}M, 0, J_{sd}^2\right\}
\label{eq:dm-1-dhad-model}
\end{equation}
and
\begin{equation}
\left[\mathbf{d}_{\eta}\times\partial_{k_a}\hat{\mathbf{d}}_{\eta}\right] = \left\{0, -\frac{\eta J_{sd}\partial_{k_a}M}{d}, 0\right\}.
\label{eq:dm-1-d-cross-dhat-model}
\end{equation}

Therefore, the first-order scalar component given in Eq.~\eqref{eq:dm-1-delta-n-solution-general} is:
\begin{equation}
n_{\eta,\mathbf{k},\omega}^{(1)} = \frac{eE^a_{\omega}\partial_{k_a} \left[f_{\eta,+} +f_{\eta,-}\right]}{\gamma_{\omega}}.
\label{eq:dm-1-delta-n-model}
\end{equation}
The first-order pseudospin component given in Eq.~\eqref{eq:dm-1-delta-s-solution-general} is:
\begin{equation}
\mathbf{s}_{\eta,\mathbf{k},\omega}^{(1)} = eE^a_{\omega}\frac{
\left(\partial_{k_a}\Delta f_{\eta}\right)}{\gamma_{\omega}} \frac{ \left\{\eta J_{sd}, 0, M\right\}}{d}
+eE^a_{\omega} \Delta f_{\eta} \frac{\gamma_{\omega} (\partial_{k_a}M) \left\{-\eta J_{sd}M, 0, J_{sd}^2 \right\}/d^3 +2\left\{0, -\eta J_{sd}(\partial_{k_a}M)/d, 0\right\}}{\gamma_{\omega}^2+4d^2}.
\label{eq:dm-1-delta-s-model-vector}
\end{equation}
The components of Eq.~\eqref{eq:dm-1-delta-s-model-vector} are:
\begin{eqnarray}
\label{eq:delta-sx-model}
s^{(1)}_{\eta,\mathbf{k},\omega;x} &=& eE^a_{\omega} \left[\frac{\eta J_{sd}}{\gamma_{\omega}d} \partial_{k_a}\Delta f_{\eta} -\Delta f_{\eta} \frac{\gamma_{\omega}\eta J_{sd}M\partial_{k_a}M}{d^3\left(\gamma_{\omega}^2+4d^2\right)}\right],\\
\label{eq:delta-sy-model}
s^{(1)}_{\eta,\mathbf{k},\omega;y} &=& -eE^a_{\omega} \Delta f_{\eta} \frac{2\eta J_{sd} \partial_{k_a}M}{d\left(\gamma_{\omega}^2 +4d^2\right)},\\
\label{eq:delta-sz-model}
s^{(1)}_{\eta,\mathbf{k},\omega;z} &=& eE^a_{\omega} \left[\frac{M}{\gamma_{\omega}d}\partial_{k_a}\Delta f_{\eta} +\Delta f_{\eta} \frac{ \gamma_{\omega} J_{sd}^2 \partial_{k_a}M}{d^3\left(\gamma_{\omega}^2 +4d^2\right)} \right].
\end{eqnarray}

\subsubsection{Second order}
\label{sec:dm-2-pwave}

The scalar second-order correction is
\begin{equation}
n^{(2)}_{\eta,\mathbf{k},\Omega} = e^2\sum_{\omega_1+\omega_2=\Omega} E^a_{\omega_1} E^b_{\omega_2} \frac{\partial_{k_a}\partial_{k_b} \left(f_{\eta,+} +f_{\eta,-} \right)}{\gamma_{\Omega}\gamma_{\omega_2}}.
\label{eq:dm-2-n-second-model}
\end{equation}
This is essentially the same as Eq.~\eqref{eq:dm-2-n-second-general}, albeit with $f_{\eta,\pm}$.

For the pseudospin part defined in Eq.~\eqref{eq:dm-2-s-second-general}, the components of $\boldsymbol{\Phi}_{\eta,b}$ are
\begin{eqnarray}
\label{eq:Phi-x-model}
\Phi^x_{\eta,b}(\mathbf{k},\omega) &=& \frac{\eta J_{sd}}{\gamma_{\omega}d} \partial_{k_b}\Delta f_{\eta} - \Delta f_{\eta} \frac{\gamma_{\omega}\eta J_{sd}M\partial_{k_b}M}{d^3\left(\gamma_{\omega}^2+4d^2\right)},\\
\label{eq:Phi-y-model}    
\Phi^y_{\eta,b}(\mathbf{k},\omega) &=& -\Delta f_{\eta} \frac{2\eta J_{sd}\partial_{k_b}M}{d\left(\gamma_{\omega}^2 +4d^2\right)},\\
\label{eq:Phi-z-model}
\Phi^z_{\eta,b}(\mathbf{k},\omega) &=& \frac{M}{\gamma_{\omega} d} \partial_{k_b}\Delta f_{\eta} + \Delta f_{\eta}\frac{\gamma_{\omega} J_{sd}^2\partial_{k_b}M}{d^3\left(\gamma_{\omega}^2 +4d^2\right)}.
\end{eqnarray}
The inverse pseudospin propagator $\mathcal{R}_{\eta,\Omega}$ given in Eq.~\eqref{eq:dm-1-R-def-general}, acting on a generic source $\mathbf{B} = \left(B_x, B_y, B_z\right)$, is:
\begin{eqnarray}
\label{eq:dm-2-R-x-model}
\mathcal{R}^x_{\eta,\Omega}[\mathbf{B}] &=& \frac{\eta J_{sd} \left( \eta J_{sd}B_x+ MB_z\right)}{d^2\gamma_{\Omega}} + \frac{\gamma_{\Omega} \left[B_x -\eta J_{sd} \left(\eta J_{sd}B_x + MB_z\right)/d^2\right] -2MB_y}{\gamma_{\Omega}^2+4d^2},\\
\label{eq:dm-2-R-y-model}
\mathcal{R}^y_{\eta,\Omega}[\mathbf{B}] &=& \frac{\gamma_{\Omega}B_y +2 \left(MB_x -\eta J_{sd}B_z \right)}{\gamma_{\Omega}^2+4d^2},\\
\label{eq:dm-2-R-z-model}
\mathcal{R}^z_{\eta,\Omega}[\mathbf{B}] &=& \frac{M\left(\eta J_{sd}B_x  + MB_z \right)}{\gamma_{\Omega}d^2} + \frac{ \gamma_{\Omega} \left[B_z - M \left( \eta J_{sd}B_x + MB_z \right)/d^2 \right] +2\eta J_{sd}B_y}{\gamma_{\Omega}^2+4d^2}.
\end{eqnarray}
The explicit expressions obtained by combining Eqs.~\eqref{eq:Phi-x-model}--\eqref{eq:Phi-z-model} and \eqref{eq:dm-2-R-x-model}--\eqref{eq:dm-2-R-z-model} are cumbersome.

\subsubsection{Orbital magnetization}
\label{sec:dm-om-pwave}

The non-Abelian off-diagonal orbital moment is represented by the following matrix in the band basis:
\begin{equation}
\hat{m}^{\rm band}_{\eta, z} = \lambda_y \frac{eJ_{sd}\alpha k_y}{4m\left(J_{sd}^2+\alpha^2k_x^2\right)} = \lambda_y \frac{eJ_{sd}\alpha k_y}{4m d^2} \equiv \tilde{m}_{\eta,z}(\mathbf{k})\lambda_y,
\label{eq:dm-m-mmat}
\end{equation}
where $\lambda_y$ acts in the band basis and we assumed $\bm{\alpha} = \alpha \hat{\mathbf{x}}$.

Therefore, the orbital magnetization in Eq.~\eqref{eq:dm-m-def} is:
\begin{equation}
\label{eq:dm-Mz}
M_z(t) = \sum_{\eta} \int \frac{d^2k}{(2\pi)^2} \, \Tr\left\{\hat m^{\rm band}_{\eta,z}(\mathbf{k})\rho^{\rm band}_{\eta,\mathbf{k}}(t) \right\} = \sum_{\eta} \int \frac{d^2k}{(2\pi)^2} \, \tilde{m}_{\eta,z}(\mathbf{k}) s^{y,{\rm band}}_{\eta,\mathbf{k}}(t).
\end{equation}
The orbital angular momentum follows from the relation $\mathbf{L} = -(1/g_L \mu_B) \mathbf{M}$.

For the continuum model given in Eq.~\eqref{ate-H-def}, $\mathbf d_{\eta}=\{\eta J_{sd},0,M(\bk)\}$ lies in the $x-z$ plane. Hence, the unitary transformation from the fixed basis to the band basis is a rotation about the $y$ axis. We obtain
\begin{equation}
s^{y,{\rm band}}_{\eta,\mathbf{k}} = \eta s^y_{\eta,\mathbf{k}},
\end{equation}
where $s^y_{\eta,\mathbf{k}}$ is the $y$ component of the density matrix obtained in the fixed $\sigma$ basis. The coefficient in the above equation is gauge-dependent and is fixed by Eq.~\eqref{eq:dm-m-unitary}. The physical observable, such as $M_z$ in Eq.~\eqref{eq:dm-Mz}, remains gauge-invariant.

Thus:
\begin{equation}
M_z^{(n)} = \sum_{\eta} \int\frac{d^2k}{(2\pi)^2}\,\tilde{m}_{\eta,z}(\mathbf{k})\eta s^{y,(n)}_{\eta,\mathbf{k}}.
\label{eq:dm-m-Mz-order-n}
\end{equation}

We now consider the orbital magnetization order by order.
At zeroth order, the pseudospin part of the density matrix is parallel to $\mathbf d_{\eta}$; see Eq.~\eqref{eq:dm-0-rho-n-s}.
Therefore:
\begin{equation}
s^{(0)}_{\eta,\mathbf{k};y}=0,
\end{equation}
and
\begin{equation}
M_z^{(0)}=0.
\label{eq:dm-m-Mz0-zero}
\end{equation}
The orbital magnetization vanishes at each $\bk$.

At first order, the correction to the $y$ component of the density matrix in the fixed $\sigma$ basis is given by Eq.~\eqref{eq:delta-sy-model}, namely:
\begin{equation}
s^{(1)}_{\eta,\mathbf{k},\omega;y} = - eE^x_{\omega} \Delta f_{\eta} \frac{ 2\eta J_{sd}\alpha}{d\left(\gamma_{\omega}^2+4d^2\right)}.
\label{eq:dm-m-sy1-alpha-y-zero}
\end{equation}
Substituting Eqs.~\eqref{eq:dm-m-mmat} and \eqref{eq:dm-m-sy1-alpha-y-zero} into Eq.~\eqref{eq:dm-m-Mz-order-n}, we obtain
\begin{equation}
M^{(1)}_{z,\omega} = - \sum_{\eta} \int\frac{d^2k}{(2\pi)^2}\, \frac{e^2J_{sd}^2\alpha^2 k_y}{2m\,d^3\left(\gamma_{\omega}^2+4d^2\right)} \Delta f_{\eta} E^x_{\omega}.
\label{eq:dm-m-Mz1-before-symmetry}
\end{equation}
The integrand is odd in $k_y$. Therefore:
\begin{equation}
M^{(1)}_{z,\omega}=0.
\label{eq:dm-m-Mz1-zero}
\end{equation}
For the momentum-resolved magnetization, an odd-in-$k_y$ component is generated by $E_x$. This is similar to the orbital Edelstein effect discussed above; see Sec.~\ref{sec:analytics-kubo-orbital} and Fig.~\ref{fig3_SM}.

At second order, the correction to the $y$ component of the density matrix in the fixed $\sigma$ basis can be written as:
\begin{equation} 
s^{y,(2)}_{\eta,\mathbf k,\Omega} = e^2 \sum_{\omega_1+\omega_2=\Omega} E^a_{\omega_1}E^b_{\omega_2} \mathcal S^y_{\eta,ab}(\mathbf k;\Omega,\omega_2), 
\label{eq:dm-m-sy2-full-decomposition} 
\end{equation} 
where $a,b=x,y$. The components of $\mathcal S^y_{\eta,ab}(\mathbf k;\Omega,\omega_2)$ are:
\begin{eqnarray} 
\label{eq:dm-m-Sy-xx-full} 
\mathcal S^y_{\eta,xx} &=& \frac{1}{D_\Omega} \left\{ 2\eta J_{sd} k_x \alpha^3 \left[ \frac{ \gamma_\Omega+2\gamma_{\omega_2} }{ d^3D_{\omega_2} } + \frac{ 8 \left( \gamma_\Omega+\gamma_{\omega_2} \right) }{ dD_{\omega_2}^2 } \right] \Delta f_\eta - \frac{ 2\eta J_{sd}\alpha }{ d } \left[ \frac{ \gamma_\Omega+\gamma_{\omega_2} }{ D_{\omega_2} } + \frac{1}{\gamma_{\omega_2}} \right] \partial_{k_x}\Delta f_\eta \right\}, \\
\label{eq:Sy-xy-full} 
\mathcal S^y_{\eta,xy} &=& - \frac{ 2\eta J_{sd}\alpha }{ dD_\Omega } \frac{ 1 }{ \gamma_{\omega_2} } \partial_{k_y}\Delta f_\eta, \\
\label{eq:dm-m-Sy-yx-full} 
\mathcal S^y_{\eta,yx} &=& - \frac{ 2\eta J_{sd}\alpha }{ dD_\Omega } \frac{ \gamma_\Omega+\gamma_{\omega_2} }{ D_{\omega_2} } \partial_{k_y}\Delta f_\eta,\\
\label{eq:dm-m-Sy-yy-full} 
\mathcal S^y_{\eta,yy} &=& 0. 
\end{eqnarray} 
Here, $D_{\omega} = \gamma_{\omega}^2+4d^2$. Equivalently, Eq.~\eqref{eq:dm-m-sy2-full-decomposition} can be rewritten as:
\begin{align} 
s^{y,(2)}_{\eta,\mathbf k,\Omega} 
&= e^2 \sum_{\omega_1+\omega_2=\Omega} E^x_{\omega_1}E^x_{\omega_2} \frac{1}{D_\Omega} \left\{ 2\eta J_{sd} k_x\alpha^3 \left[ \frac{ \gamma_\Omega+2\gamma_{\omega_2} }{ d^3D_{\omega_2} } + \frac{ 8 \left( \gamma_\Omega+\gamma_{\omega_2} \right) }{ dD_{\omega_2}^2 } \right] \Delta f_\eta \right. \nonumber\\ 
&\left. - \frac{ 2\eta J_{sd}\alpha }{ d } \left[ \frac{ \gamma_\Omega+\gamma_{\omega_2} }{ D_{\omega_2} } + \frac{1}{\gamma_{\omega_2}} \right] \partial_{k_x}\Delta f_\eta \right\} \nonumber\\ 
&- e^2 \sum_{\omega_1+\omega_2=\Omega} \frac{ 2\eta J_{sd}\alpha }{ dD_\Omega } \left[ E^y_{\omega_1}E^x_{\omega_2} \frac{ \gamma_\Omega+\gamma_{\omega_2} }{ D_{\omega_2} } + E^x_{\omega_1}E^y_{\omega_2} \frac{1}{\gamma_{\omega_2}} \right] \partial_{k_y}\Delta f_\eta . 
\label{eq:dm-m-sy2-full-alpha-y-zero} 
\end{align} 
The first term in Eq.~\eqref{eq:dm-m-sy2-full-alpha-y-zero} is responsible for the $xx$ component of the second-order orbital-magnetization response tensor. It is even in $k_y$ and odd in $k_x$. The second term in Eq.~\eqref{eq:dm-m-sy2-full-alpha-y-zero} provides the mixed $xy+yx$ contribution. It is odd in $k_y$ through $\partial_{k_y}\Delta f_\eta$ and is even in $k_x$. 

The second-order orbital magnetization is therefore determined by:
\begin{align}
M^{(2)}_{z,\Omega} &= \sum_{\eta} \int \frac{d^2k}{(2\pi)^2} \, \tilde{m}_{\eta,z}(\mathbf{k})  s^{(2), {\rm band}}_{\eta,\mathbf{k},\Omega;y} \nonumber\\
&= \sum_{\eta} \int \frac{d^2k}{(2\pi)^2} \,
\frac{e^3J_{sd}\alpha k_y}{4m d^2} \sum_{\omega_1+\omega_2=\Omega} E^x_{\omega_1}E^x_{\omega_2} \frac{1}{D_\Omega} \left\{ 2 J_{sd} k_x\alpha^3 \left[ \frac{ \gamma_\Omega+2\gamma_{\omega_2} }{ d^3D_{\omega_2} } + \frac{ 8 \left( \gamma_\Omega+\gamma_{\omega_2} \right) }{ dD_{\omega_2}^2 } \right] \Delta f_\eta \right. \nonumber\\ 
&\left. - \frac{ 2J_{sd}\alpha }{d} \left[ \frac{ \gamma_\Omega+\gamma_{\omega_2} }{ D_{\omega_2} } + \frac{1}{\gamma_{\omega_2}} \right] \partial_{k_x}\Delta f_\eta \right\}
\nonumber\\
& - \sum_{\eta} \int \frac{d^2k}{(2\pi)^2} \, \frac{e^3J_{sd}^2\alpha^2 k_y}{2m\,d^3 D_{\Omega}} 
\sum_{\omega_1+\omega_2=\Omega} \left[ E^y_{\omega_1}E^x_{\omega_2} \frac{ \gamma_{\Omega} + \gamma_{\omega_2}}{D_{\omega_2}} +E^x_{\omega_1}E^y_{\omega_2}\frac{1}{\gamma_{\omega_2}}\right] \partial_{k_y}\Delta f_{\eta}.
\label{eq:dm-m-Mz2-alpha-y-zero}
\end{align}
The derivative $\partial_{k_y}\Delta f_{\eta}$ is odd in $k_y$:
\begin{equation}
\partial_{k_y}\Delta f_{\eta} = \frac{k_y}{m} \left(f'_{\eta,+} -f'_{\eta,-}\right).
\end{equation}
Therefore, only the last term can contribute to the total magnetization because it is even in $k_y$. The first term is odd in both $k_y$ and $k_x$; hence, it vanishes after integration over momentum. 

The dc response corresponds to $\Omega=\omega_1=\omega_2=0$. The second-order rectified response also has $\Omega=0$, but $\omega_1=-\omega_2$. For a monochromatic field, the sum $\sum_{\omega_1+\omega_2=\Omega}$ contains two terms: $\omega_1=-\omega_2 =\omega$ and $\omega_1=-\omega_2 =-\omega$.
Finally, the second harmonic generation response has $\Omega=\pm 2\omega$ with $\omega_1=\omega_2=\pm \omega$.

Combining the results in Eqs.~\eqref{eq:dm-m-Mz0-zero}, \eqref{eq:dm-m-Mz1-zero}, and \eqref{eq:dm-m-Mz2-alpha-y-zero}, we write
\begin{equation}
M_z \approx M_z^{(0)} +M_z^{(1)} +M_z^{(2)}.
\end{equation}
It follows that $M_z$ requires both $E_x$ and $E_y$. A field $\mathbf{E}\parallel \hat{\mathbf{x}}$ does not contribute to the integrated orbital magnetization for the symmetric model with $\alpha_y=0$, while for $\mathbf{E}\parallel \hat{\mathbf{y}}$, $M_z$ vanishes because $\partial_{k_y}M(\bk)=0$.

Using Eq.~\eqref{eq:dm-m-dimless}, we define the following dimensionless momentum-resolved magnetization:
\begin{equation}
\label{eq:dm-m-dimless-m}
\frac{m_z(\Omega, \bk)}{m_{z,0}} = \chi^{(1)}_{za}(\Omega; \bk)\mathcal{E}^a_{\Omega} +\sum_{\omega_1+\omega_2 = \Omega} \chi^{(2)}_{zab}(\Omega,\omega_2;\bk)\mathcal{E}^a_{\omega_1}\mathcal{E}^b_{\omega_2} +\ldots,
\end{equation}
where $\mathcal{E}^a_{\omega} = E^a_{\omega}/E_0$ and $\chi^{(n)}_{za_1\ldots a_n}$ is the momentum-resolved susceptibility tensor, which is essentially the integrand of the susceptibility tensor; see also Eq.~\eqref{eq:dm-m-dimless} for the definitions of the characteristic variables. For example, according to Eq.~\eqref{eq:dm-m-Mz1-before-symmetry},
\begin{equation}
\chi^{(1)}_{zx}(\Omega; \bk) = - \sum_{\eta} \frac{2J_{sd}^3\alpha^2k^{\star} k_y }{d^3\left(\gamma_{\Omega}^2+4d^2\right)} \Delta f_{\eta},
\quad \chi^{(1)}_{zx}(\Omega; \bk) =0.
\end{equation}

In calculating the second-order susceptibility $\chi^{(2)}_{zab}$, we focus on the rectified component with $\Omega=0$ and $\omega_1=-\omega_2=\omega$. The corresponding magnetization is given by:
\begin{eqnarray}
\label{eq:dm-m-dimless-m-2}
\frac{m_{z}^{(2)}(\Omega=0, \bk)}{m_{z,0}} &=& \chi^{(2)}_{zab}(\Omega=0,-\omega; \bk)\mathcal{E}^a_{\omega}\mathcal{E}^b_{-\omega} +\chi^{(2)}_{zab}(\Omega=0,\omega; \bk)\mathcal{E}^a_{-\omega}\mathcal{E}^b_{\omega}
=2\Re\left\{\chi^{(2)}_{zab}(\Omega=0,-\omega; \bk)\mathcal{E}^a_{\omega}\mathcal{E}^b_{-\omega} \right\} \nonumber\\
&=& 2\Re\chi^{(2)}_{zab}(\Omega=0,-\omega; \bk) \Re\left\{\mathcal{E}^a_{\omega}\mathcal{E}^b_{-\omega}\right\}
-2\Im\chi^{(2)}_{zab}(\Omega=0,-\omega;\bk) \Im\left\{\mathcal{E}^a_{\omega}\mathcal{E}^b_{-\omega}\right\}.
\end{eqnarray}

We use the following model parameters unless otherwise stated:
\begin{equation}
\label{eq:dm-m-dimless-numvars}
\mu=0, \quad \alpha = 0.5\,\alpha_{0}, \quad \tau = 100\,\tau_0, \quad T=0.1\,J_{sd}.
\end{equation}

\begin{figure}[t]
   \centering
   \includegraphics[width=0.99\textwidth]{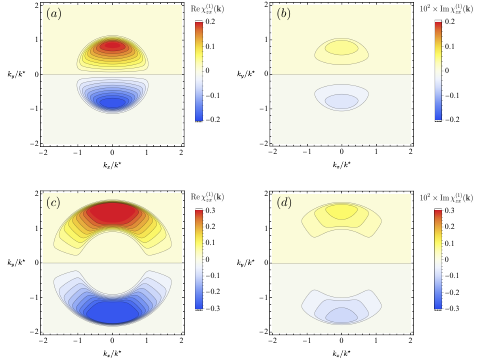}
   \caption{
   The dimensionless first-order momentum-resolved orbital-magnetization susceptibility $\chi^{(1)}_{za}(\omega;\bk)$ as a function of momentum at fixed $\omega=0.5\,J_{sd}$. Left and right columns correspond to the real and imaginary parts of the susceptibility, respectively. Top and bottom rows show the results for $\mu=0$ and $\mu=2\,J_{sd}$, respectively. We used the characteristic parameters in Eq.~\eqref{eq:dm-m-dimless} and the numerical parameters in Eq.~\eqref{eq:dm-m-dimless-numvars}.}
   \label{fig:dm-m-chi1}
\end{figure}

\begin{figure}[t]
   \centering
   \includegraphics[width=0.99\textwidth]{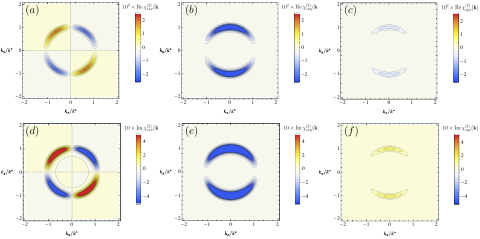}
   \caption{The dimensionless second-order momentum-resolved orbital-magnetization susceptibility $\chi^{(2)}_{zab}(\Omega=0,-\omega;\bk)$ as a function of momentum at $\omega=0.5\,J_{sd}$ and $\mu=0$. Top and bottom panels correspond to the real and imaginary parts of the susceptibility. Note that $\chi^{(2)}_{zyy}(\Omega=0,-\omega;\bk) = 0$. We used the characteristic parameters in Eq.~\eqref{eq:dm-m-dimless} and the numerical parameters in Eq.~\eqref{eq:dm-m-dimless-numvars}.}
   \label{fig:dm-m-chi2-mu=0}
\end{figure}

\begin{figure}[t]
   \centering
   \includegraphics[width=0.99\textwidth]{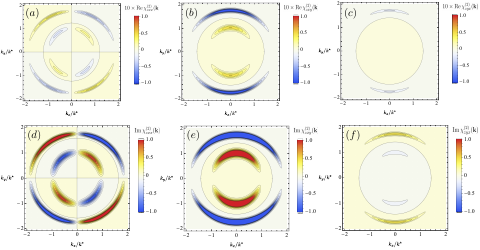}
   \caption{The dimensionless second-order momentum-resolved orbital-magnetization susceptibility $\chi^{(2)}_{zab}(\Omega=0,-\omega;\bk)$ as a function of momentum at $\omega=0.5\,J_{sd}$ and $\mu=2\,J_{sd}$. Top and bottom panels correspond to the real and imaginary parts of the susceptibility. Note that $\chi^{(2)}_{zyy}(\Omega=0,-\omega;\bk) = 0$. We used the characteristic parameters in Eq.~\eqref{eq:dm-m-dimless} and the numerical parameters in Eq.~\eqref{eq:dm-m-dimless-numvars}.}
   \label{fig:dm-m-chi2-mu=2}
\end{figure}

We show the momentum-resolved first-order orbital magnetization in Fig.~\ref{fig:dm-m-chi1} and its second-order counterpart in Figs.~\ref{fig:dm-m-chi2-mu=0} and \ref{fig:dm-m-chi2-mu=2}. Temperature $T$ determines the sharpness of the outer boundary in Fig.~\ref{fig:dm-m-chi1} and the width of the ``pockets'' in Figs.~\ref{fig:dm-m-chi2-mu=0} and \ref{fig:dm-m-chi2-mu=2}.

In agreement with our analytical result in Eq.~\eqref{eq:dm-m-Mz1-before-symmetry}, the first-order momentum-resolved orbital magnetization is odd in momentum; see Fig.~\ref{fig:dm-m-chi1}.

Numerical results in Figs.~\ref{fig:dm-m-chi2-mu=0} and \ref{fig:dm-m-chi2-mu=2} confirm that the static orbital-magnetization texture can be generated by second-order processes. However, the total orbital magnetization requires a light field with components both parallel and transverse to $\bm{\alpha}$. Circularly polarized light provides another example of light capable of generating rectified orbital magnetization. The momentum-resolved static orbital-magnetization texture is also interesting: it shows lobes with $d$-wave symmetry, whose number is doubled at $\mu=2\,J_{sd}$ when both upper and lower bands contribute; see Fig.~\ref{fig:dm-m-chi2-mu=2}.

\begin{figure}[t]
   \centering
   \includegraphics[width=0.99\textwidth]{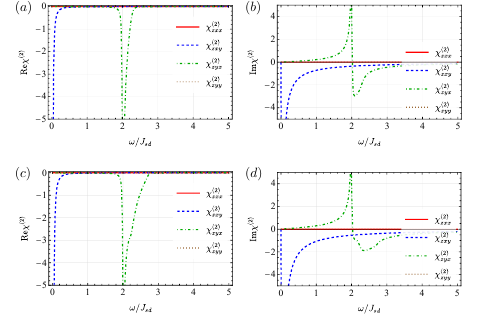}   
   \caption{The dimensionless momentum-integrated orbital-magnetization susceptibility $\chi^{(2)}_{zab}(\Omega=0,-\omega)$ as a function of $\omega$. Left and right columns correspond to the real and imaginary parts of the susceptibility, respectively. Top and bottom rows show the results for $\mu=0$ and $\mu=2\,J_{sd}$, respectively. We used the numerical parameters in Eq.~\eqref{eq:dm-m-dimless-numvars}.}
   \label{fig:dm-m-chi2-int}
\end{figure}

Unlike the momentum-resolved orbital-magnetization susceptibility $\chi^{(2)}_{zab}(\Omega=0,-\omega;\bk)$ shown in Figs.~\ref{fig:dm-m-chi2-mu=0} and \ref{fig:dm-m-chi2-mu=2}, its momentum-integrated counterpart shown in Fig.~\ref{fig:dm-m-chi2-int} does not show any qualitative dependence on $\mu$.

\bibliography{Library-short}